# Generalized Methods and Solvers for Noise Removal from Piecewise Constant Signals

Max A. Little (max.little@physics.ox.ac.uk), Nick S. Jones (nick.jones@physics.ox.ac.uk)

Department of Physics and Oxford Centre for Integrative Systems Biology, University of Oxford, UK

22 December 2010

## Abstract

Removing noise from signals which are piecewise constant (PWC) is a challenging signal processing problem that arises in many practical scientific and engineering contexts. For example, in exploration geosciences, noisy drill hole records must be separated into constant stratigraphic zones, and in biophysics, the jumps between states and dwells of a molecular structure need to be determined from noisy fluorescence microscopy signals. This problem is one for which conventional linear signal processing methods are fundamentally unsuited. A wide range of PWC denoising methods exists, including total variation regularization, mean shift clustering, stepwise jump placement, running median filtering, convex clustering shrinkage, bilateral filtering, wavelet shrinkage and hidden Markov models. This paper builds on results from the image processing community to show that the majority of these algorithms, and more proposed in the wider literature, are each associated with a special case of a generalized functional, that, when minimized, solves the PWC denoising problem. We show how the minimizer can be obtained by a range of computational solver algorithms, including stepwise jump placement, quadratic or linear programming, finite differences with and without adaptive step size, iterated running medians, least angle regression, piecewise-linear regularization path following, or coordinate descent. Using this generalized functional, we introduce several novel PWC denoising methods, which, for example, combine the global behaviour of mean shift clustering with the local smoothing of total variation diffusion, and show example solver algorithms for these new methods. Head-to-head comparisons between these methods are performed on synthetic data, revealing that our new methods have a useful role to play. Finally, overlaps between the generalized methods of this paper and others such as wavelet shrinkage, hidden Markov models, and piecewise smooth filtering are touched on.

## 1. Introduction

*Piecewise constant (PWC) signals* exhibit flat regions with a finite number of abrupt jumps that are instantaneous, or effectively instantaneous because the transitions occur in between sampling intervals. These signals occur in many contexts, including bioinformatics (Snijders et al., 2001), astrophysics (OLoughlin, 1997), geophysics (Mehta et al., 1990), molecular biosciences (Sowa et al., 2005) and digital imagery (Chan and Shen, 2005). Figure 1 shows examples of signals that could fit this description that are apparently contaminated by significant noise. Often, we are interested in recovering the PWC signal from this noise, using some kind of digital filtering technique.

Because such signals arise in a great many scientific and engineering disciplines, this noise filtering problem turns out to be of enduring interest. However, it goes under a confusing array of names. An abrupt jump can be called a *shift*, *edge*, *step*, *change*, *change point*, or less commonly, *singularity* or *transition* (sometimes combined, e.g. *step change*), and to emphasise that this jump is instantaneous, it can occasionally also be *sharp*, *fast* or *abrupt*. The constant regions are often also called *levels*. Bearing in mind that finding the location of the jumps usually allows estimation of the level of the flat regions, the filtering process itself (usually *smoothing*) can also be called *detection* or *approximation*, and less commonly *classification*, *segmentation, finding* or *localization*.

Statisticians have long been interested in this and related problems. Some of the earliest attempts to solve the related *change point detection* problem arose in the 1950's for *statistical process control* in manufacturing (Page, 1955), which began a series of statistical contributions that continues to this day, see for example (Pawlak et al., 2004). The *running median filter* was introduced in the 1970's (Tukey, 1977) as a proposed improvement to *running mean filtering*, bringing *robust statistical estimation* theory to bear on this problem. Following this, robust statistics features heavily in a diverse range of approaches reported in the statistics (Fried, 2007), signal processing (Elad, 2002; Dong et al., 2007) and applied mathematics literature (Gather et al., 2006).

The PWC with noise model is also important for digital images, because edges, corresponding to abrupt image intensity jumps in a scan line, are highly salient features (Marr and Hildreth, 1980). Therefore, noise removal from PWC signals is of critical importance to *digital image processing*, and a very rich source of contributions to the PWC filtering problem have been devised in the image signal processing community, such as *mathematical morphology* (Serra, 1982), *nonlinear diffusion filtering* (Perona and Malik, 1990), *total variation denoising* (Rudin et al., 1992) and related approaches, developed through the 1970's to this day. These efforts established strong connections with, and assimilated some of the earlier work on, robust filtering (Elad, 2002; Mrazek et al., 2006). The fact that *piecewise Lipschitz functions* are good models for PWC signals implies that they have a parsimonious representation in a *wavelet basis* (Mallat, 2009), and wavelets for PWC denoising were introduced in the 1990's (Mallat and Hwang, 1992).

In apparent isolation from the image processing and statistics communities, other disciplines have described alternative algorithms. Beginning in the 1970's, exploration geophysicists devised a number of novel PWC denoising algorithms, including *stepwise jump placement* (Gill, 1970) – apparently reinvented almost 40 years later by biophysicists (Kerssemakers et al., 2006). In the 1980's *hidden Markov models* (Godfrey et al., 1980) were introduced by geophysicists, with biophysics following this trend in the 1990's (Chung et al., 1990). Neuroscientists described novel nonlinear filters that attempt to circumvent the edge smoothing limitations of running mean filtering around the same time (Chung and Kennedy, 1991).

Superficially, this problem does not appear to be particularly difficult, and so it is reasonable to ask why it still deserves attention. To answer this from a signal processing perspective, abrupt jumps pose a fundamental challenge for *conventional linear methods*, e.g. finite impulse response, infinite impulse response or fast Fourier transform-based filtering. In the Fourier basis, PWC signals *converge slowly*: that is, the magnitudes of Fourier coefficients decrease much slower with increasing frequency than the coefficients for continuous functions (Mallat, 2009). Signal recovery requires removing the noise, and conventional linear methods typically achieve this by *low-pass filtering*, that is, by removal of the high frequency detail in the signal. This is effective if the signal to be recovered is sufficiently smooth. But PWC signals are not smooth, and low-pass filtering of PWC signals typically introduces large, spurious oscillations near the jumps known as *Gibb's phenomena* (Mallat, 2009). The noise and the PWC signal *overlap substantially in the Fourier basis* and so cannot be separated by any basic approach that reduces the magnitude of some Fourier coefficients, which is how conventional low-pass noise removal works. This typical inadequacy of conventional linear filtering is illustrated in Figure 2. Therefore, we usually need to invoke *nonlinear* or *non-Gaussian* techniques (that is, where the variables in the problem are not assumed to be normally distributed) in order to achieve effective performance in this digital filtering task. The nonlinearity or non-Gaussianity of these techniques makes them harder to understand than linear techniques, and, as such, there is still much to discover about the PWC denoising problem, and it remains a topic of theoretical interest.

The literature on this topic is fragmented across statistics, applied mathematics, signal and image processing, information theory and specialist scientific and engineering domains. Whilst relationships between many of algorithms discussed here have been established in the image processing and statistics communities – such as the connections between nonlinear diffusion, robust filtering, total variation denoising, mean shift clustering and wavelets (Candes and Guo, 2002; Elad, 2002; Steidl et al., 2004; Chan and Shen, 2005; Mrazek et al., 2006; Arias-Castro and Donoho, 2009) – here we identify some broader principles at work:

1. The problem of PWC denoising is fruitfully understood as either *piecewise constant smoothing*, or as *level-set recovery* owing to the fact that typically, there will be either only a few isolated jumps in the signal, or just a few, isolated levels. The piecewise constant view naturally suggests methods that fit *0-degree (constant) splines* to the noisy signal and which typical find the *jump locations* that determine the levels. By contrast, the level-set view suggests *clustering* methods that attempt to find the levels and thus determine the location of the jumps,
2. Building on work from the image processing literature, all the methods we study here are associated with special cases of a generalized, functional equation, with the choice of terms in this functional determining the specifics of each PWC method. A few, general "component" functions are assembled into the terms that go to make up this functional. We show here that this functional is broadly applicable to a wide set of methods proposed across the disciplines,
3. All these methods function, either explicitly by the action of the solver, or implicitly by nature of the generalized functional, by application of a *sample distance reduction principle*: to minimize the sum in the functional, the absolute differences between some samples in the input signal has to reduce sufficiently to produce solutions that have what we call the *PWC property*. A solution with this property has a parsimonious representation as a constant spline or level-set,
4. All the PWC methods we study here attempt to minimize the generalized functional obtained using some kind of *solver*. Although, as presented in the literature, these solvers are all seemingly very different, we show that these are in fact special cases of a handful of very general concepts, and we identify the conditions under which each type of solver can be applied more generically.

These findings provide us with some structural insights about existing methods and their relationships that we explore in this paper, and allow us to develop a number of novel PWC denoising techniques, and some new solvers, that blend the relative merits of existing methods in useful ways. The detailed nature of the extensive ground work at the start of this paper is necessary to make it clear how the novel methods we propose in later sections are both relevant, useful and solvable in practice.

A summary of the paper is as follows. Section 2 motivates and formalizes the spline and level-set models for discrete-time PWC signals. Section 3 introduces the generalized functional that connects all the methods in this paper, and describes how this functional can be built from component functions. It introduces the sample distance reduction principle. It shows how existing PWC denoising algorithms are associated with special cases of this functional. Section 4 discusses general classes of solvers that minimize the generalized functional, and some new observations about existing PWC denoising methods that arise when considering the properties of these solvers. Synthesising the knowledge from these earlier sections, Section 5 then goes on to motivate and devise new PWC denoising methods and solvers. Section 6 compares the numerical results of two challenging PWC denoising tasks, and discusses the accuracy of methods and efficiency of different solvers. Finally, Section 7 summarises the findings of the paper and connects to other approaches, including wavelets, HMMs, piecewise smooth filters and nonlinear diffusion PDEs, and mentions possible directions for future research.

## 2. Piecewise constant signals as splines and level-sets

In this paper, we wish to recover an $N$ sample PWC signal $m_i \in \mathbb{R}$, for $i = 1, 2 \ldots N$. We assume that the discrete-time signal is obtained by sampling of the continuous-time signal $m(t)$, $t \in [t_1, t_N]$ (note that the use of "time" here simply stands in for the fact that the signal is just a set of values ordered by the index $i$ or $t$, and we will often suppress the index for notational clarity). The observed signal is corrupted by an additive noise random process $e_i \in \mathbb{R}$, i.e. $x = m + e$.

PWC signals consist of two fundamental pieces of information: the levels (the values of the samples in constant regions), and the boundaries of those regions (the locations of the jumps). A common theme in this paper is the distinction between (a) PWC signals described by the locations of the jumps, which in turn determine the levels according to the specifics of the noise removal method, and (b) signals described by the values of the levels, which then determine the location of the jumps through the properties of the method.

By way of motivating the jump interpretation, we consider Steidl *et al.* (2006) showing that the widely used *total variation regularization* PWC denoising method has, as solutions, a set of discrete-time, *constant 0-degree splines*, where the location of the spline knots is determined by the regularization parameter $\gamma$ and the input data $x$. This result provides the first intuitive model for PWC signals as constructed from constant splines, and PWC denoising as a *spline interpolation problem*. The spline model is usually a compact one because it is generally the case that the PWC signal to be recovered has only a small number of discontinuities relative to the length of the signal, that is, only a few jumps occurring between indices $i$ and $i+1$ where $m_i \neq m_{i+1}$. The $M$ jumps in the signal occur at the *spline knots* with locations $\{r_1, r_2, \ldots r_{M+1}\}$ (together with the "boundary knots" $r_0 = 1$ and $r_{M+1} = N + 1$ for completeness). The PWC signal is reconstructed from the values of the constant levels $\{l_1, l_2, \ldots l_{M+1}\}$ and the knot locations, e.g. $m_i = l_j$ for $r_{j-1} \leq i < r_j$, where $j = 1, 2 \ldots M + 1$.

Alternatively, one can view the problem of PWC denoising as a *clustering problem*, classically solved using techniques such as *mean shift* or *K-means clustering* (Cheng, 1995). In this context, it is natural to apply the *level-set* model, and indeed, this may sometimes be more useful (and more compact) than the spline description (Chan and Shen, 2005). The level-set for the value $l \in \Omega$ ($\Omega$ refers to the set of all unique values in the PWC signal), is the set of indices corresponding to $l$, $\Gamma(l) = \{i : m_i = l\}$. The complete level-set over all values of the PWC signal $\Gamma$ is formed from the union of these level-sets, which also makes up the complete index set, $\Gamma = \bigcup_{l \in \Omega} \Gamma(l) = \{1, 2 \ldots N\}$. The level-sets form a partition of the index set, so that $\Gamma(l_A) \cap \Gamma(l_B) = \emptyset$ for all $l_A \neq l_B$ where $l_A, l_B \in \Omega$. A succinct representation of each level-set can be constructed using only the left and right boundary indices of each contiguously numbered range of indices that make up each level-set. The spline and level-set descriptions are, of course, readily interchangeable using appropriate transformations.

Since this paper is concerned with discrete-time signals only, the definition of a PWC signal used in this paper is that they have a *simple representation* as either 0-degree splines or as level-sets. By simple, we mean that the number of jumps is small compared to the number of samples, $M/N \ll 1$, or, that the number of unique levels is small compared to the number of samples $|\Omega|/N \ll 1$. If a signal satisfies either condition we say that it has the *PWC property*.

## 3. A generalized functional for PWC denoising

As discussed in the introduction, all the PWC denoising methods investigated in this paper are associated with special cases of the following general functional equation:

$$H[m] = \sum_{i=1}^{N} \sum_{j=1}^{N} \Lambda(x_i - m_j, m_i - m_j, x_i - x_j, i - j) \tag{3.1}$$

Here $x$ is the input signal of length $N$, and $m$ is the output of the noise removal algorithm, of length $N$. This functional combines *difference* functions into *kernels*, and *losses*. See Tables 1 and 2 and the next section for details. In practice, useful kernel and loss functions for PWC denoising are typically of the form described in the tables. A large number of existing methods can be expressed as special cases of the resulting functional assembled from these functional components (Table 1). Various *solvers* can be used to minimize this functional to obtain the output $m$, these are listed in Table 3.

### 3.1 Differences, kernels and losses

As described in Table 1, the basis of the unification of these methods into a single functional equation, is the quantification of the *differences between all pairs* of input $x$ and output samples $m$, and their indices $i, j$ (Table 1a). In the statistical literature, the generalized functional (3.1) would typically be derived from specification of *likelihood* and *prior* distributions, where the likelihood would involve terms in $x_i - m_j$ and the prior involve functions of $m_i - m_j$. A minimizer for the functional would be a *regularized maximum likelihood* or *maximum a-posteriori estimator*. In this paper, we will therefore describe terms in $x_i - m_j$ as *likelihood* terms, and terms in $m_i - m_j$ as *regularization* terms.

Using these differences, *loss functions* (Table 1c) and *kernels* (Table 1b) are constructed. By kernels, here we simply mean non-negative functions of absolute difference (we call this *distance*), which are usually symmetric. The loss functions are non-negative functions of distances. We define two different kinds of losses: *simple* losses that increase with distance, and *composite losses* that are only increasing with distance over a certain range of the distance. The derivative of the loss function: the *influence function* (a term borrowed from the robust statistics literature) plays an important role in some iterative algorithms for minimizing the functional (in particular, see Section 4.5, on finite differences below). With composite loss functions, the influence function is seen to be a product of an associated kernel term that represents the *magnitude* of the gradient of the loss, and a term that represents the *direction* of the gradient of the loss. In this paper, we will focus on simple symmetric distance functions. The three cases we will focus on the *non-zero count* $p = 0$ defining $|d|^0$, which is zero if $d$ is zero, and one otherwise. The case $p = 1$ corresponds to the *absolute* distance, and $p = 2$ corresponds to the *square* distance $|d|^2/2$.

We distinguish between differences in the *values* of input and output samples, $x_i - m_j$, $m_i - m_j$ and $x_i - x_j$, and the difference between the *sequence* of samples $i - j$. Thus, a kernel based on differences between pairs of variables $x, m$ we call a *value kernel*, to distinguish it from a kernel based on $i - j$ which we call a *sequence kernel*. We make further distinctions between *hard* and *soft kernels*. Hard kernels are non-zero for some range of distances, and outside this range, they are zero. Soft kernels take non-zero values for all values of the distance. We also describe the trivial kernel that is 1 for all values of distance as the *global kernel*. When used as a sequence kernel the global kernel means that all pairwise terms enter into the sum, and when used as a value kernel it implies that all differences in value are weighted equally. All other kernels are therefore *local* kernels. The special local sequence kernels $I(d = 1)$ and $I(d = 0)$ isolate only *adjacent* terms in the generalized functional sum, and terms that are *aligned* to the same index value, respectively (where $I(S)$ is an indicator function which takes a value of 1 if $S$ is true and zero otherwise).

The loss functions are assembled into the function $\Lambda$ in (3.1) that quantifies the loss incurred by every difference. Summation of $\Lambda$ over all pairs of indices in the input and output signals leads to the functional $H[m]$ to be minimized with respect to the output $m$.

### 3.2 The sample distance reduction principle

The generalized presentation of the PWC denoising methods in this paper allows us to see that the basic operation of these methods is to reduce the distance between samples in the input signal. In this section we give a non-rigorous explanation for this behaviour. As the simplest example, consider $\Lambda = |m_i - m_j|^p / p$; for $p \geq 1$ this leads to a convex functional that has the optimum, constant solution $m_i = c$ (this can be shown by differentiating $H$ with respect to each $m_i$ and setting each equation to zero). Throughout the paper we use the notation $m^k$ to denote the output signal obtained at iteration $k$ of a solver (we thus have a mixed notation in which the context defines the interpretation of $m$: it can either be the unknown PWC signal we are trying to estimate or represents our current best estimate). Our solvers would typically be initialised with $m^0 = x$ and then successive attempts at solutions, $m^k$, are conditional on past attempts). We expect good iterative solvers initialized with $m^0 = x$ to reduce the distance between input samples in successive iterations, the natural termination of this process being the constant solution $m_i = c$. This occurs with the simple loss $|m_i - m_j|^p / p$ that increases with increasing difference, and minimizing the total sum of losses requires that the differences must be reduced in absolute value.

Of course, this trivial constant solution is no use in practice. One way in which this trivial solution is avoided is by *regularization*: for the purpose of illustration, consider the functional arising from $\Lambda = (1/p)|x_i - m_j|^p I(i - j = 0) + \gamma/p|m_i - m_j|^p$ for $p \geq 1$ (see Table 2). The resulting functional has the property that when the regularization parameter $\gamma = 0$ the optimal solution is $m = x$; but as $\gamma \to \infty$, the second term dominates, forcing the samples in the output signal to collapse onto a single constant. A useful PWC output consisting of several different levels might lie between these two extremes.

The trivial constant solution is also avoided by the introduction of kernels. Consider, for example, the soft-mean shift functional $\Lambda = 1 - \exp(-\beta|m_i - m_j|^p/p)/\beta$ for $p \geq 1$ (See Table 2), and an iterative solver initialized with $m^0 = x$. With this modification to the simple loss function (Table 1c), the loss attached to distances between samples does not increase strongly with increasing differences: beyond a certain distance, the loss remains effectively unchanged. Thus, in minimizing the total sum of losses in the functional, some pairs of samples are forced closer together, whereas others are free to become further apart. Those that are constrained eventually collapse onto a few levels. Therefore, a minimum of the functional is often a useful PWC solution. Note that the trivial constant solution is a minimizer, but because the functional is not convex, a non-trivial PWC solution is usually reached first by a gradient descent solver.

Sequence kernels allow the distance reduction to become localised in index. For the *diffusion filter* $\Lambda = |m_i - m_j|^p I(i - j = 1)$ with $m^0 = x$ and $p \leq 1$, only samples that are adjacent to each other must become closer to each other under minimization of the functional (see Section 4.3). The difference between samples that are not adjacent is irrelevant. Locally constant runs of similar values can therefore emerge to produce a PWC output. Note that here, for the case $p = 2$, the only possible PWC output is the trivial constant output because the diffusion is then linear.

Kernels applied to differences of the input samples alone can also prevent the output from collapsing down onto a single constant. For example, by modifying the simple loss (Table 1c) with the hard kernel (Table 1b) applied to the input differences, as in $\Lambda = (1/p)|m_i - m_j|^p I(|x_i - x_j|^p/p \leq W)$, $p \geq 1$, with solver initialization $m^0 = x$, only those samples in the output signal that *have the same index* as samples in the input signal that are close in value, end up making a contribution to the sum in the functional. Because of this, minimizing the functional requires only that the distance between those samples in the output signal must be reduced, the rest are unconstrained. Therefore, the outputs that minimize this (convex) functional can include ones that consist of more than one level.

### 3.3 Existing methods in the generalized functional form

*Diffusion filtering-type methods*

These methods, with $\Lambda = (1/q)|x_i - m_j|^q I(i - j = 0) + \gamma|m_i - m_j|^p I(i - j = 1)$, can be understood as combining sequentially aligned likelihood terms with adjacent regularization terms (see Section 3.1), using simple losses, with the regularization parameter $\gamma$. We mention the case $q = p = 2$ for completeness: this can be solved using a *(cyclic) running weighted mean filter* or using Fourier filtering (see Section 4.3). It is, however, of no practical use in PWC denoising because it is purely quadratic, and hence has a linear filtering operation as solver, a situation discussed in the introduction. Of more value is the case where $q = 2$ and $p = 1$: this is *total variation regularization* (Rudin et al., 1992). Where $q = 2$ and $p = 0$, we obtain many *jump placement* methods that have been proposed in the scientific and engineering literature (Gill, 1970; Kerssemakers et al., 2006; Kalafut and Visscher, 2008). The corresponding diffusion filtering methods, that are not constrained by the input signal (but that typically have the signal as the initial condition of an iterative solver: $m^0 = x$), are obtained when the likelihood term is removed, e.g. with $\Lambda = (1/p)|m_i - m_j|^p I(i - j = 1)$.

*Convex clustering shrinkage*

This clustering method has $\Lambda = (1/2)|x_i - m_j|^2 I(i - j = 0) + \gamma|m_i - m_j|$, and combines aligned differences in the likelihood term with a global regularization term with regularization parameter $\gamma$. It uses only simple losses. The likelihood term uses the square loss, whereas the regularization term has absolute value loss (Pelckmans et al., 2005).

*Mean shift clustering-type methods*

This class of methods uses global likelihoods or regularizers, where the losses (Table 1c) are associated with hard, local value kernels (Table 1b). For $\Lambda = \min(|m_i - m_j|, W)$, coupled with an adaptive step-size finite difference solver, we have *mean shift clustering*, and with $\Lambda = \min(|x_i - m_j|, W)$ we obtain a clustering method that has important similarities to *K-means clustering*, we will call this *likelihood mean shift* (Fukunaga and Hostetler, 1975; Cheng, 1995), also see Section 4.6. Since these methods use composite losses as defined in Table 1c, differences between samples have to be small in order to make a difference to the value of the functional. Hence, samples that start off close under some iterative solver initialised with $m^0 = x$, will become closer under iteration of the solver, this induces the "clustering" effect of these methods (see Section 4.6 for further details).

*Bilateral filtering-type methods*

These methods exploit soft value kernels, and hard sequence kernels in the regularization term, and have $\Lambda = [1 - \exp(-\beta|m_i - m_j|)/\beta]I(|i - j| \leq W)$. One way of describing these methods is that they are similar to mean shift clustering with soft value kernels, but combined with sequentially local, hard kernels (Mrazek et al., 2006). They therefore inherit some of the clustering effect of mean shift clustering, but also the effect of clustering due to sequence locality.

## 4. Solvers for the generalized functional and some new observations for existing methods

We distinguish two broad classes of solvers for the generalized functional: (a) those that directly minimize the functional, and (b) those that solve the *descent ordinary differential equations* (ODEs) obtained by differentiating the functional with respect to $m$. In category (a) we find *greedy* methods that attempt to fit a 0-degree spline to the noisy signal, convex optimization methods including *linear* and *quadratic programming*, *coordinate descent*, *subgradient* and many others. In category (b) we find a very large number of techniques that can be identified as *numerical methods* for the (simultaneous) *initial value problem* we obtain by differentiating the functional with respect to the output signal $m_i$. The goal of this section is to discuss these solvers in the context of important PWC denoising methods that have found frequent use in practice.

Here we expand upon the descent ODEs in a special case that is important for those solvers in category (b). A minimum of the generalized functional is obtained at $\partial H/\partial m_i = 0$ for each $i = 1,2 \ldots N$ (which parallels the first-order optimality condition in variational calculus). It will not be possible in general to solve this resulting set of equations analytically, so one approach is to start with a "guess" solution $m = a$ and to evolve this trial solution in the direction that lowers the value of $H$ the most, until the solution stops changing at a minimum of the functional. This is the idea behind the (steepest) descent ODEs defined as $dm_i/d\eta = -\partial H/\partial m_i$, with the initial conditions $m_i(0) = a_i$. The solution depends on the solver parameter $\eta$. Many of the algorithms we describe in this paper can be written in the form $\Lambda = F(x_i - m_j)\kappa_1(i - j) + \gamma G(m_i - m_j)\kappa_2(i - j)$ where $F, G$ are loss functions, $\kappa_{1,2}$ are any sequence kernels, and $\gamma$ is the regularization parameter, and the steepest descent ODEs are then:

$$\frac{dm_i}{d\eta}(\eta) = -\frac{\partial H}{\partial m_i} = -\sum_{j=1}^{N} F'\left(x_j - m_i(\eta)\right)\kappa_1(i-j) - \gamma \sum_{j=1}^{N} G'\left(m_i(\eta) - m_j(\eta)\right)\kappa_2(i-j) \quad (4.1)$$

Here the dependence of the outputs on the solver parameter $\eta$ has been made explicit, but we will usually suppress this for clarity. Typically, it is arranged such that, when $\eta = 0$, $m = x$, and $x$ is often used as the initial condition for these ODEs. As the ODEs are evolved forward in $\eta$, the output $m$ becomes closer to having the PWC property on each iteration.

### 4.1 Stepwise jump placement

A conceptually simple and commonly proposed algorithm for directly minimizing $H[m]$ is *stepwise jump placement* that starts with a spline with no knots as a trial solution and then introduces them to the spline one at a time (Gill, 1970; Kerssemakers et al., 2006; Kalafut and Visscher, 2008). The location of each new knot is determined by *greedy search*, that is, by a systematic scan through all locations $i = 1,2 \ldots N$, finding the location that reduces the functional the most at each iteration. If the iteration stops after a few knots, this ensures that the solutions satisfy the PWC property. At iteration $k$ we denote the spline knot locations as $\{r_1, r_2, \ldots r_k\}$. Then the values of the constant levels $\{l_1, l_2, \ldots l_{k+1}\}$ are determined that minimize the generalized functional given these fixed knot indices. Stepwise jump placement methods typically define a functional of the form:

$$H[m] = f\left(\sum_{i=1}^{N}\sum_{j=1}^{N}(1/2)|x_i - m_j|^2 I(i-j=0)\right) + g\left(\sum_{i=1}^{N}\sum_{j=1}^{N}|m_i - m_j|^0 I(i-j=1)\right) \tag{4.2}$$

where $f, g$ are strictly increasing functions – and since they are increasing, this functional has the same minimizer as the functional obtained from $\Lambda = (1/2)|x_i - m_j|^2 I(i-j=0) + \lambda|m_i - m_j|^0 I(i-j=1)$, with a regularization parameter $\lambda > 0$ that is determined by either the properties of the input signal or the choice of the number of jumps. In particular, the method of Kalafut and Visscher (2008) has $f(s) = N\log(s)$ and $g(s) = \log(N)s$. Since the number of jumps is fixed at each iteration, the optimum levels in the spline fit are just the mean of the samples $x$ for each level:

$$l_j = \frac{1}{r_j - r_{j-1}} \sum_{i=r_{j-1}}^{(r_j)-1} x_i \tag{4.3}$$

for $j = 1, 2 \ldots k+1$. Only the likelihood term must be evaluated to perform the greedy scan for the index of each new knot at iteration $k+1$. Given the functional above, it can be that no new knot index can be found that reduces $H[m]$ below the previous iteration; this is used as a criteria to terminate the placement of new knots (Gill, 1970; Kalafut and Visscher, 2008). Stopping after a predetermined number of jumps have been placed (Gill, 1970), or determining a peak in the ratio of the likelihood term to the likelihood evaluated using a "counter-fit" (Kerssemakers et al., 2006), similar in spirit to the *F-ratio statistic in analysis of variance*, are two other suggested termination criteria.

### 4.2 Linear and quadratic programming

For purely *convex problems* (that is, problems where the loss functions are all convex in $m$), the unique minimizer for $H[m]$ can be found using standard techniques from convex optimization (Boyd and Vandenberghe, 2004). In particular, both total variation regularization (Rudin et al., 1992) and convex clustering shrinkage (Pelckmans et al., 2005) can be transformed into a quadratic program (quadratic problem with linear inequality constraints), which can be solved by *interior-point* techniques. Fast, specialized *primal-dual* interior-point methods for total variation regularization have been developed recently (Kim et al., 2009). The scope for linear programs is very wide, it applies to loss functions such as the loss based on the absolute distance, but also for asymmetric *quantile loss* functions such as $L(d) = [q - I(d < 0)]d$, where $q$ is the appropriate quantile $q \in [0,1]$. Quantiles are minimizers for these asymmetric losses, the median being the special, symmetric case (Koenker, 2005), and these losses would be useful if it is expected that the noise distribution has asymmetric outliers.

### 4.3 Analytic solutions to the descent ODEs

In general, all useful PWC methods have functionals that cannot be minimized analytically; it is informative for the flow of this paper, however, to study a functional that can be solved analytically, even though it is not a useful in practice. For the special case of simple square loss functions, minimization of the functional can be carried out directly using matrix arithmetic. We start by considering *linear diffusion filtering*:

$$\Lambda = (1/2)|m_i - m_j|^2 I(i-j=1) \tag{4.4}$$

The associated initial value descent ODEs are:

$$\frac{dm_i}{d\eta} = -\frac{\partial H}{\partial m_i} = m_{i+1} - 2m_i + m_{i-1} \tag{4.5}$$

with $m(0) = x$, the boundary cases defined by $m_i \equiv 0$ for $i < 1$ and $i > N$. We can write this in matrix form as $dm/d\eta = Am$ where $A$ is the *system matrix* with $-2$ on the main diagonal, and $+1$ on the diagonals above and below the main diagonal. This can be understood as a *semi-discrete heat equation*, with the right hand side being a discrete approximation to the Laplacian. This set of homogeneous, linear, constant coefficient ODEs can be solved exactly by finding the eigenvalues $\lambda$ and eigenvectors of the system matrix $A$ which are:

$$\lambda_i = -2 + 2\cos\left(\frac{i\pi}{N+1}\right), V_{ij} = \sin\left(\frac{ij\pi}{N+1}\right), \quad i, j = 1, 2 \ldots N \tag{4.6}$$

The matrix of eigenvectors $V$ is orthogonal, and can be made orthonormal without loss of generality. This matrix is then unitary so $V = V^T = V^{-1}$, and the solution is written explicitly in terms of the eigenvectors:

$$m(\eta) = V \begin{bmatrix} c_1 \exp(\lambda_1 \eta) \\ \vdots \\ c_N \exp(\lambda_N \eta) \end{bmatrix} \qquad (4.7)$$

The $N$ constants of integration $c$ are determined by the initial condition $m(0) = x$, by calculating $c = Vx$. This matrix operation can, in fact, be seen to be the *discrete sine Fourier transform* of the input signal, so the constants are Fourier coefficients of the expansion of the solution in the sine basis, and the solution is merely the inverse discrete sine transform of the discrete sine Fourier domain representation of the input signal, where each frequency component is scaled by $\exp(\lambda_i \gamma)$. Since the eigenvalues are always negative, the contribution of these frequency components in the solution decay with increasing $\eta$, tending to zero as $\eta \to \infty$. This confirms, by a different route, that the solution can only be entirely constant when all samples are zero. Additionally, $\lambda_{i+1} < \lambda_i$ for all $i = 1, 2 \ldots N$ so that high frequency components decay more quickly with increasing $\eta$ than lower frequency components. Therefore, high frequency fluctuations due to noise are quickly smoothed away, and slowly-varying frequency components remain.

We will now make a connection to the *weighted running mean filter*, a ubiquitous linear smoothing technique. The linearity and translation invariance with respect to $\eta$ of this system allows the solution to be written in terms of a (circular) convolution with the Green's function (impulse response in the signal processing literature). Using the special initial condition $m_i(0) = 1$ for $i = \lfloor N/2 \rfloor$ and $m_i(0) = 0$ otherwise, the Green's function is:

$$h = V \left[ (Vm(0)) \circ \begin{bmatrix} \exp(\lambda_1 \Delta \eta) \\ \vdots \\ \exp(\lambda_N \Delta \eta) \end{bmatrix} \right] \qquad (4.8)$$

for a particular $\Delta \eta > 0$ (here $\circ$ denotes the entrywise product). Because multiplication of the frequency components is equivalent to convolution in the domain $i$, we can now write the solution as:

$$m_i(\Delta \eta) = h \star x_i = \sum_{j=-N/2}^{N/2-1} h_{(j-1) \bmod N + 1} x_{(i+j-1) \bmod N + 1} \qquad (4.9)$$

where $\star$ indicates circular convolution. The Green's function $h$ is of the form of a Gaussian "pulse" centred in the middle of the signal. Iterating the convolution $k$-times, $\star_k$, gives the solution at multiples of $\Delta \eta$, i.e. $m(k \Delta \eta) = h \star_k x$. For small $\Delta \eta$, the Gaussian pulse has small effective width and so the Green's function, centered around the Gaussian pulse, can be truncated to produce an (iterated) *weighted running mean filter* with short window length $(2W + 1) < N$:

$$m_i^{k+1} = \sum_{j=-W}^{W} h_j m_{i-j}^k \qquad (4.10)$$

with $m^0 = x$ and the $2W + 1$ weights, obtained by centering and truncating the Green's function, are normalized $\sum_{j=-W}^{W} h_j = 1$. At the boundaries we define $m_i \equiv 0$ for $i < 1$ and $i > N$. The smoothing behaviour of this linear filter is useful for noise removal, but, as discussed in the introduction, since jumps in PWC signals also have significant frequency contributions at the scale of noise fluctuations, these are smoothed away simultaneously. Thus, the smoothing functional obtained by the square regularization loss is of little practical value in PWC denoising applications, despite the tantalizing availability of an exact analytic minimizer and its practical implementation as a simple running weighted mean filter.

### 4.4 Iterated running median filter

While it was seen above that the iterated running (weighted) mean filter is of no essential value in noise removal from PWC signals due to its linearity, the nonlinear *iterated running median filter* has been proposed instead. This finds the median (rather than the mean) of the samples in a window of length $2W + 1$ that slides over the signal:

$$m_i^{k+1} = \text{median}(m_{i-W}^k, \ldots, m_{i+W}^k) = \underset{\mu \in \mathbb{R}}{\text{argmin}} \sum_{j=-W}^{W} |m_{i+j}^k - \mu| \qquad (4.11)$$

with $m^0 = x$, and the boundaries are defined through $m_i \equiv 0$ for $i < 1$ and $i > N$. The above minimization expresses the idea that the median is the constant $\mu$ that minimizes the total absolute deviations from $\mu$ of the samples in each window. This contrasts with the (equal weighted) running mean filter which minimizes the total *squared* deviations instead. It is well-known that the running median filter does not smooth away edges as dramatically as the running mean filter under conditions of low noise spread (Justusson, 1981; Arias-Castro and

Donoho, 2009), and therefore this filter has value as a method for PWC denoising in a limited range of applications.

Iterated median filtering has some value as a method for PWC denoising, so it is interesting to ask how it is related to other methods in this paper. We observe here a connection between *total variation diffusion filtering* and the iterated median filter. We prove in the appendix that applying the median filter with window size $2W + 1 = 3$ to a signal cannot increase the *total variation* of the signal, e.g. $TV[m^{k+1}] \leq TV[m^k]$, where $TV[m] = \sum_{i=1}^{N-1}|m_{i+1} - m_i|$. If we consider a numerical solver for the total variation diffusion ODEs obtained from the generalized functional with $\Lambda = |m_i - m_j|I(i - j = 1)$:

$$\frac{dm_i}{d\eta} = \text{sgn}(m_i - m_{i+1}) - \text{sgn}(m_i - m_{i-1}) \tag{4.12}$$

with the initial condition $m(0) = x$, this solver must also reduce the total variation on each iteration (because it is an integrator that lowers the total variation functional at each iteration). The window length 3 iterated median filter differs from such an integrator because every iterated median filter converges on a *root signal* that depends on $x$, that is, a signal that is fixed under the iteration of the filter (Arce, 2005). Therefore, unlike the solution to the total variation diffusion ODEs (that eventually leads to a constant signal with zero total variation) this iterated median filter cannot remove all jumps for all signals $x$, and so it does not necessarily reduce the total variation to zero. Determining the knots in the spline representation is not a simple matter for the iterated median filter. After convergence, whether the solutions have the PWC property depends upon the initial conditions, and the number of iterations to reach convergence.

### 4.5 Finite differences

Few other solvers have such widespread applicability as numerical methods for the descent ODEs (4.1). For example, in Section 4.6 we will see that many important PWC clustering algorithms can be derived as special cases of such numerical methods. Initial value problems such as (4.1) can be approximately integrated using any of a wide range of numerical methods, including *Euler* (forward) *finite differences* (Mrazek et al., 2006):

$$m_i^{k+1} = m_i^k - \Delta\eta \sum_{j=1}^{N} F'(m_i^k - x_j)\kappa_1(i - j) - \gamma\Delta\eta \sum_{j=1}^{N} G'(m_i^k - m_j^k)\kappa_2(i - j) \tag{4.13}$$

where $\Delta\eta$ is the step size, together with initial condition $m_i^0 = a_i$, a set of constants.

This is accurate to first order in the step size. Higher order accurate integrators could be used instead if required. In the special case where all the loss functions are convex and differentiable, this method must converge on the unique minimizer for $H[m]$. If any one of the loss functions is not differentiable everywhere, then convergence is not guaranteed, but achieving a good approximation to the minimizer may still be possible, particularly if the loss function is non-differentiable at only a small set of isolated points. If the loss functions are not convex but are differentiable, then convergence to a minimizer for the functional is guaranteed; but this may not be the minimizer that leads to the *smallest possible* value for the functional. Without differentiability, then convergence cannot be guaranteed either. For non-convex losses, one useful heuristic to gain confidence that a proposed solution found using finite differences is the minimizer associated with the smallest possible value for the functional is to restart the iteration several times from randomized starting conditions and iterate until convergence (or approximate convergence). One can then take the solution with the smallest value of the functional from these (approximately) converged solutions.

### 4.6 Finite differences with adaptive step sizes

In this section we will obtain many standard clustering algorithms as special cases of the finite differences introduced above. For the Euler forward finite difference solver, the fixed step size $\Delta\eta$ can be replaced with an adaptive step size. This trick can be used to derive *mean shift*, and the *soft* version of this method, as well as the *bilateral filter* (Mrazek et al., 2006), but it can be used more generally. We note here that the popular *K-means* method is conceptually extremely similar although not a direct special case of the functional (3.1). In this section, we show how to derive a method we call *likelihood mean shift* (see Table 2) that *is* a relevant special case of the functional (3.1).

As discussed earlier, if the loss function is composite (Table 1c), then the influence function is the product of a kernel and a direction term (Cheng, 1995). In particular, for the local, hard loss functions $\min(|d|, W)$ and $\min(|d|^2/2, W)$, the influence functions are $I(|d| \leq W) \times \text{sgn}(d)$ and $I(|d|^2/2 \leq W) \times d$, so in the latter case, the kernel is the hard window of size $W$, and the direction term is just the difference $d$.

With composite square loss functions, such as $\min(|d|^2/2, W)$, and by (4.13), the Euler finite difference formula can be:

$$m_i^{k+1} = m_i^k - \Delta\eta \sum_{j=1}^{N} I\left(\left|m_i^k - x_j\right|^2/2 \le W\right)\left(m_i^k - x_j\right)\kappa_S(i-j)$$
$$- \gamma\Delta\eta \sum_{j=1}^{N} I\left(\left|m_i^k - m_j^k\right|^2/2 \le W\right)\left(m_i^k - m_j^k\right)\kappa_S(i-j) \tag{4.14}$$

where $\kappa_S$ is any sequence kernel (here, for simplicity, we have shown the case where the form of the kernels used in the likelihood and regularization terms are the same, but they need not be in general). Now, we set an appropriate adaptive step size:

$$\Delta\eta_i = \left[\sum_{j=1}^{N} I\left(\left|m_i^k - x_j\right|^2/2 \le W\right)\kappa_S(i-j) + \sum_{j=1}^{K} I\left(\left|m_i^k - m_j^k\right|^2/2 \le W\right)\kappa_S(i-j)\right]^{-1} \tag{4.15}$$

ensuring steps become larger in flatter regions. Classical mean shift (Section 3.3 and Table 2) uses the hard local, square loss function; the sequence kernel is global, so the finite difference formula becomes:

$$m_i^{k+1} = m_i^k - \Delta\eta \sum_{j=1}^{N} I\left(\left|m_i^k - m_j^k\right|^2/2 \le W\right)\left(m_i^k - m_j^k\right) \tag{4.16}$$

Replacing the step size with the adaptive quantity $\Delta\eta_i = \left(\sum_{j=1}^{N} I\left(\left|m_i^k - m_j^k\right|^2/2 \le W\right)\right)^{-1}$, after some algebra we get:

$$m_i^{k+1} = \frac{\sum_{j=1}^{N} I\left(\left|m_i^k - m_j^k\right|^2/2 \le W\right) m_j^k}{\sum_{j=1}^{N} I\left(\left|m_i^k - m_j^k\right|^2/2 \le W\right)} \tag{4.17}$$

which is the classical mean shift algorithm that replaces each output sample value with the mean of all those within a distance $W$. What we are calling *likelihood mean shift* (Section 3.3 and Table 2), has, similar to mean shift the adaptive step-size, $\Delta\eta_i = \left(\sum_{j=1}^{N} I\left(\left|m_i^k - m_j^k\right|^2/2 \le W\right)\right)^{-1}$ leading to the iteration:

$$m_i^{k+1} = \frac{\sum_{j=1}^{N} I\left(\left|m_i^k - m_j^k\right|^2/2 \le W\right) x_j}{\sum_{j=1}^{N} I\left(\left|m_i^k - m_j^k\right|^2/2 \le W\right)} \tag{4.18}$$

that replaces each cluster centroid $m_i, i = 1 \ldots N$ with the mean of all the *input samples* within a distance $W$. Soft versions of both algorithms are obtained by using the soft kernel instead of the hard kernel.

Up until now it has been assumed that that for each sample value at $i$, $x_i$, there is a corresponding estimate for the PWC signal $m_i$; in this case $1 \le i \le N$ is acting as an index for "time" for both input and output signals. For our particular discussion of $K$-means below it is necessary to allow that the index of $m_i$ need not be a proxy for time but instead indexes each distinct level in the PWC output signal: there might be $K$ distinct levels in the PWC output signal and it is possible that $K < N$. Deriving the classical *K-means algorithm* – requires the construction of the value kernel:

$$\kappa_C(m_i, x_j) = I\left(m_i = \underset{1 < \alpha < K}{\operatorname{argmin}} |m_\alpha - x_j|\right) \tag{4.19}$$

which is the indicator function of whether the cluster centroid $m$ is the closest to the input sample $x$. Then the iteration:

$$m_i^{k+1} = \frac{\sum_{j=1}^{N} \kappa_C(m_i^k, x_j) x_j}{\sum_{j=1}^{N} \kappa_C(m_i^k, x_j)} \tag{4.20}$$

can be seen to replace the cluster centroids with the mean of all samples that are closer to it than to any other centroid. Cheng (1995) shows that $\kappa_C(m_i, x_j)$ can be obtained as the limiting case of the smooth function:

$$\frac{\exp\left(-\beta(m_i - x_j)^2/2\right)}{\sum_{p=1}^{K} \exp\left(-\beta(m_p - x_j)^2/2\right)} \to \kappa_C(m_i, x_j) \tag{4.21}$$

when $\beta \to \infty$. Indeed, for finite $\beta$, this yields the soft $K$-means algorithm. However, as we discussed above (Section 3.3), there are two reasons why the classical $K$-means algorithm departs from the generalized functional (3.1) in this paper. The first is because the number of distinct output samples in the $K$-means algorithm is $K \leq N$, $m_i$ for $i = 1,2 \ldots K$. However, if there are many less than $N$ levels in a PWC signal, the $K$-means solver typically merges the input samples down onto this small number of unique output values. The second departure is that the kernel $\kappa_C$ cannot be obtained directly from the particular form of the generalized functional (3.1), because each term $\Lambda$ must then be a function of differences of *all* samples in $m$ and $x$, not just differences of samples indexed by the pair $i,j$. However, $K$-means is an important PWC method and it is conceptually very similar to mean shift. In fact, the really critical difference is that the $K$-means algorithm iterates on the likelihood difference $x_i - m_j$, whereas mean shift iterates on the regularization difference $m_i - m_j$ (compare (4.18) with (4.20)) This is our reason for calling the clustering method based on the likelihood $x_i - m_j$ the likelihood mean shift method.

The *bilateral filter* (Section 3.3 and Table 2) combines the hard local sequence kernel $I(|i - j| \leq W)$ and the soft loss term $1 - \exp\left(-\beta|m_i - m_j|^2/2\right)/\beta$ and this leads to the following finite difference update:

$$m_i^{k+1} = m_i^k - \Delta\eta \sum_{j=1}^{N} \exp\left(-\beta|m_i^k - m_j^k|^2/2\right)(m_i^k - m_j^k)I(|i - j| \leq W) \tag{4.22}$$

Inserting the adaptive step size $\Delta\eta_i = \left(\sum_{j=1}^{N} \exp\left(-\beta|m_i^k - m_j^k|^2/2\right)I(|i - j| \leq W)\right)^{-1}$ obtains the bilateral filter formula (Mrazek et al., 2006):

$$m_i^{k+1} = \frac{\sum_{j=1}^{N} \exp\left(-\beta|m_i^k - m_j^k|^2/2\right)I(|i - j| \leq W)m_j^k}{\sum_{j=1}^{N} \exp\left(-\beta|m_i^k - m_j^k|^2/2\right)I(|i - j| \leq W)} \tag{4.23}$$

See also Elad (2002) for a very instructive alternative derivation involving Jacobi solvers for the equivalent matrix algebra formulation.

This section has shown how adapting the step-size of the Euler integrator leads to a number of well-known clustering algorithms for appropriate combinations of loss functions. The dynamics of the evolving solution can be understood in terms of the level-set model. For mean shift clustering, initially, $m^0 = x$, and (assuming noise), each $m_i^0$ will typically have a unique value, so every level-set contains one entry (which is just the index for each sample), $\Gamma(m_i) = i$. As the iterations proceed, Cheng (1995) shows that if $W$ is sufficiently large that the support of the hard value kernel covers more than one sample of the initial signal, these samples within the support will be drawn together until they merge onto a single value after a finite number of iterations. After merging, they always take on the same value under further iterations. Therefore, after merging, there will be a decreased number of unique values in $m$, and fewer unique level-sets, that will consist of an increased number of indices. Groups of merged samples will themselves merge into larger groups under subsequent iterations, until a fixed point is reached at which no more changes to $m^k$ occur under subsequent iterations. Therefore, after convergence, depending on the initial signal and the width of the kernel, there will typically only be a few level-sets that will consist of a large number of indices each, and the level-set description will be very compact.

In the case of $K$-means clustering, there are $K$ values in the PWC signal output $m^k$ and at each step, every level-set at iteration $k$ is obtained by evaluating the indicator kernel $\kappa_C$ for every $i = 1,2 \ldots N$: $\Gamma(m_i^k) = \{j \in 1,2 \ldots N : \kappa_C(m_i^k, x_j) = 1\}$. Note that it is possible for two of the levels to merge with each other, in which case the associated level-sets are also merged. After a few iterations, $K$-means converges on a fixed point where there are no more changes to $m^k$ (Cheng, 1995). Soft kernel versions of $K$-means and mean shift have similar merging behaviour under iteration, except the order of the merging (that is which sets of indices are merged together at each iteration) will depend in a more complex way upon the initial signal and the kernel parameter $\beta$.

Bilateral filtering can be seen as soft mean shift, but with the addition of a hard sequential window. Therefore it inherits similar merging and convergence behaviour under iteration. However, for samples to merge, they must both be close in value *and* temporally separated by at most $W$ samples (whereas for mean shift, they need only be close in value). The additional constraint of temporal locality implies that each merge does not necessarily assimilate large groups of indices, and the level-set description is not typically as compact as with mean shift.

### 4.7 Piecewise linear path following
For nearly all useful functionals of the form (3.1), analytical solutions are unobtainable. However, it turns out that there are some important special cases which for which a minimizer can be obtained with algorithms that might be described as *semi-analytic*, and we describe them in this section. For useful PWC denoising, it is common that the right hand side of the descent ODE system is discontinuous, which poses a challenge for conventional numerical techniques such as finite differences. However, it has been shown that if the likelihood term is convex and *piecewise quadratic* (that is, constructed of piecewise polynomials of order at most two), and the regularization term has convex loss functions that are *piecewise linear*, then the solution to the descent ODEs is continuous and constructed of piecewise linear segments (Rosset and Zhu, 2007). Formally, there is a set of *L regularization points* $0 = \gamma_0 < \gamma_1 < \cdots < \gamma_L = \infty$ and a corresponding set of $N$-element *gradient* vectors $\epsilon^0, \epsilon^1 \ldots \epsilon^L$, in terms of which the full *regularization path*, that is, the set of all solutions obtained by varying a regularization parameter $\gamma \geq 0$, can be expressed. We can write this as:

$$m(\gamma) = m(\gamma_j) + (\gamma - \gamma_j)\epsilon^j, \qquad \gamma_j \leq \gamma \leq \gamma_{j+1} \tag{4.24}$$

for all $j = 0,1 \ldots L - 1$. PWC denoising algorithms that have this piecewise linear regularization path property include *total variation regularization* and *convex clustering shrinkage* (Pelckmans et al., 2005). The values of the regularization points and the gradient vectors can be found using a general solver proposed by Rosset and Zhu (2007), but specialized algorithms exist for total variation regularization; one finding the path in "forward" sequence of increasing $\gamma$ (Hofling, 2009), and the other, by expressing the convex functional in terms of the *convex dual variables* (Boyd and Vandenberghe, 2004), obtains the same path in reverse for decreasing $\gamma$ (Tibshirani and Taylor, 2010).

Total variation regularization has been the subject of intensive study since its introduction (Rudin et al., 1992). Strong and Chan (2003) show that a step of height $h$ and width $w$ in an otherwise zero signal is decreased in height by $2\gamma/w$, and is "flattened" when $2\gamma/w \geq h$. Fundamentally, these findings can be explained by the sample reduction principle: the form of the regularization term acts to linearly decrease the absolute difference in value between adjacent samples $m_i(\gamma)$ and $m_{i-1}(\gamma)$ as $\gamma$ increases (a process known as *shrinkage* in the statistics literature), and once adjacent samples eventually coincide for one of the regularization points $\gamma_j$, they share the same value for all $\gamma \geq \gamma_j$. Thus, pairs of samples can be viewed as merging together (a process known as *fusing*) to form a new partition of the index set, consisting of subsets of indices in consecutive sequences with no gaps. We will see illustrations of this behaviour later when we examine the iteration paths of other solvers as well.

Initially, at $\gamma = \gamma_0$, this partition is the trivial one where each subset of the index set contains a single index. Subsets of indices in the current partition assimilate their neighbouring subsets as $\gamma$ increases, until the partition consists of just one subset containing all the indices at $\gamma = \gamma_{L-1}$, and this is also where $m_i = E[x]$. Thus, total variation regularization recruits samples into constant "runs" of increasing length as $\gamma$ increases.

This offers another intuitive explanation for why constant splines afford a compact understanding of the output of total variation regularization. For the backward path following solver (Tibshirani and Taylor, 2010) that begins at the regularization point $\gamma_{L-1}$, the spline consists of no jumps, and only the boundary knots $r_0 = 1$, $r_1 = N + 1$ and one level $l_1 = E[x]$. As the path is followed backward to the next regularization point $\gamma_{L-2}$, the spline is split with a new knot at location $i$ and one new level $l_2$ is added, so that the spline is described by the set of knots $\{r_0 = 1, r_1 = i, r_2 = N + 1\}$ and levels $\{l_1, l_2\}$. The solver continues adding knots at each regularization point until there are $N$ levels and $N + 1$ knots. The forward path following algorithm starts at this condition and merges levels by deleting knots at each regularization point.

Piecewise linear path following requires the computation of the regularization points $\gamma_j$, and it is possible to directly compute the maximum useful value of the regularization parameter where all the output samples are fused together (Kim et al., 2009):

$$\gamma_{L-1} = \|(DD^T)^{-1}Dx\|_\infty \tag{4.25}$$

where $\|.\|_\infty$ is the elementwise vector maximum, and $D$ is the $N \times N$ first difference matrix:

$$D = \begin{bmatrix} 1 & -1 & & & \\ & 1 & -1 & & \\ & & \ddots & \ddots & \\ & & & 1 & -1 \\ & & & & 1 \end{bmatrix} \tag{4.26}$$

Furthermore, knowing that a step of height $h$ and unit width is flattened when $\gamma \geq h/2$, allows us to suggest an estimate for the *minimum* useful value that is just larger than the noise spread: if the noise is Gaussian with standard deviation $\sigma$, then setting $\gamma \geq 2\sigma$ will remove 99% of the noise. Therefore, the useful range of the regularization parameter for PWC denoising can be estimated as $2\sigma \leq \gamma \leq \gamma_{L-1}$.

### 4.8 Other solvers

The descent ODEs define an initial value problem that is a standard topic in the numerical analysis of nonlinear differential equations, and there exists a substantial literature on numerical integration of these equations (Iserles, 2009). These include the finite difference methods discussed above, but also *predictor-corrector* and higher order methods such as *Runge-Kutta*, *multistep* integrators, and *collocation*. The cost of higher accuracy with high order integrators is that an increased number of evaluations of the right hand side of the descent ODEs are required per step. However, the main departure of this problem from classical initial value problems is the existence of discontinuities in the right hand side of the descent ODE system that arise when the loss functions are not differentiable everywhere, and most of the useful loss functions for PWC denoising methods are non-differentiable. As a solution, *flux* and *slope-limiters* have been applied to total variation regularization in the past (Rudin et al., 1992). We also mention here the very interesting matrix algebra interpretation of PWC denoising methods that opens up the possibility of using solvers designed for numerical matrix algebra including the *Jacobi* and *Gauss-Seidel* algorithms, and variants such as *successive over-relaxation* (Elad, 2002).

## 5. New methods and solvers for PWC denoising

Having introduced the components, the generalized functional and solver algorithms for existing methods, in this section we investigate how some of these existing concepts can be generalized. There is more than one potential starting point for this. One approach is to ask about the range of validity of their associated solvers: what properties must the functional satisfy to allow this solver to be applied? Another approach is to attempt to synthesise new functionals that are "hybrids" of existing methods, leading to new methods that have their own merit as PWC denoising methods. We will start by seeing how the very simplest stepwise jump placement solvers can be generalized (Section 5.1). We then discuss the connection between total variation regularization and *regression splines*, and in doing so motivate a novel *coordinate descent* method in (Section 5.2). By considering a generalization of total variation regularization, we will give a novel convex method that can handle statistical *outliers* in the noise, and can be solved using off-the-shelf linear programming algorithms (Section 5.3). Next, in addressing an important limitation of convex clustering shrinkage (see Section 3.3 and Table 2), we will motivate a weighting trick that not only improves the usefulness of convex clustering shrinkage, but also leads to a novel version of mean shift clustering that provides a fundamentally new clustering method and associated solver algorithm (Section 5.4). Finally, by exposing some of the limitations of total variation diffusion and mean shift clustering, we develop a hybrid method with improved performance, and derive a new solver algorithm for it (Section 5.5).

### 5.1 Jump penalization and robust jump penalization

Stepwise jump placement methods can ensure that the solutions have the PWC property, which makes it interesting to ask whether the idea can be generalized. The conceptual simplicity of the stepwise jump placement solver algorithm is frustrated if the regularization term depends on the knot locations, as in the case of total variation regularization where the regularization term involves the absolute value of adjacent differences, or where minimizing the likelihood term given the fixed knot configuration is not straightforward or requires considerable computational effort. Thus, the greatest appeal of stepwise jump placement algorithms is as a minimizer for functionals that combine the non-zero count regularization term with adjacent sequence kernel, $|m_i - m_j|^0 I(i - j = 1)$, but, more generally, likelihood terms such as $(1/p)|x_i - m_j|^p I(i - j = 0)$, where $p \geq 1$. We can therefore suggest novel *jump penalization* methods:

$$\Lambda = (1/p)|x_i - m_j|^p I(i - j = 0) + \gamma |m_i - m_j|^0 I(i - j = 1) \tag{5.1}$$

for $p \geq 1$ and freely chosen regularization parameter $\gamma \geq 0$. For $p = 2$, the mean formula (4.3) applies when calculating the levels of the spline fit, whereas for $p = 1$ the median formula is required to calculate the levels instead:

$$l_j = \text{median}\left(x_{r_{j-1}}, x_{r_{j-1}+1}, \ldots x_{(r_j)-1}\right) \tag{5.2}$$

(Recall that $r_j$ is the time index of the $j^{th}$ knot of the spline). From a statistical point of view, this jump penalization method with $p = 1$ is valuable where the noise distribution is symmetric and heavy-tailed, because in this situation the mean will be heavily influenced by outliers, but the median is robust to these large deviations. The functional is non-convex and non-differentiable, and thus not amenable to methods such as linear or quadratic programming, and will pose non-convergence challenges for numerical methods for the associated initial value problem. However, the greedy search used in stepwise jump placement requires reconstructing the spline fit for each putative new jump location and this is not necessarily computationally efficient.

In the relevant literature (Gill, 1970; Kerssemakers et al., 2006; Kalafut and Visscher, 2008), we have only found the idea that stepwise jump placement proceeds with introducing new knots until a termination criteria is reached. However, this stepwise jump placement strategy has the disadvantage that the minimizer that leads to the smallest possible value of the functional might only be achievable by stepwise *removal* of jumps. Therefore it may be necessary to place a jump at every location, and perform iterative *jump removal* to attempt to lower the functional. Similarly, because the non-zero count loss is non-convex, the functional is not convex either, and there may be another solution that lowers the functional further. Therefore, in some circumstances, alternating between iterative knot placement and iterative knot removal may succeed in finding a better solution than either placement or removal alone. In fact, minimizing the functional is a combinatorial optimization problem, because the number of knots is an integer quantity. Therefore, it can be addressed by the wide array of techniques that have been developed for such problems (Papadimitriou and Steiglitz, 1998).

The jump penalization methods introduced above have another useful interpretation where the PWC signal represents a discrete-time stochastic process that can have both positive and negative jumps of any height. The count number of a *Poisson process* is an important special case of this where the jumps are all of the same height and positive only: then the time interval between jumps is exponentially distributed. In that case, the probability of obtaining a jump in any one discrete-time sampling interval is just $\rho = \tau/\mu$, where $\tau$ is the sampling interval and $\mu$ is the mean time between jumps. In the corresponding discrete-time setting, the number of jumps is a random variable that is Bernoulli distributed with parameter $\rho$. Then the appropriate choice of regularization parameter is $\gamma = \log((1-\rho)/\rho)$. At one extreme, when $\rho = 1/2$, that is, a jump is exactly as likely as no jump in any one sampling interval, this factor is zero, so the number of jumps plays no role in the minimizer of the functional, which is just the input signal $x$. At the other extreme, when $\rho \to 0$, the mean time between jumps becomes infinite, so a jump in any interval becomes improbable, and $\gamma \to \infty$. This forces the number of jumps to zero when minimizing the functional.

### 5.2 Regression splines and coordinate descent

In this section we demonstrate the intimate connection between *total variation regularization*, which is of major importance in PWC denoising applications and *spline regression*, and how a simple new solver can be applied to find the solution. For the special case of total variation regularization, for which $\Lambda = (1/2)|x_i - m_j|^2 I(i - j = 0) + \gamma |m_i - m_j| I(i - j = 1)$, the functional becomes:

$$H[m] = \frac{1}{2} \sum_{i=1}^{N} (m_i - x_j)^2 + \gamma \|Dm\|_1 \tag{5.3}$$

where $\|.\|_1$ is the (entrywise) vector 1-norm, and $D$ is the $N \times N$ first difference matrix as defined in (4.26). This is shown to be equivalent to the following functional (Kim et al., 2009):

$$H[\mu] = \frac{1}{2} \|S\mu - x\|_2^2 + \gamma \|\mu\|_1 \tag{5.4}$$

where $\|\cdot\|_p$ is the (entrywise) vector $p$-norm, to be minimized over the $N$ new variables $\mu_i$ (these new variables are spline coefficients related to the original variables $m$, see below). The $N \times N$ matrix $S = (D^{-1})^T$ has the form:

$$S = \begin{bmatrix} 1 & & & & \\ 1 & 1 & & & \\ 1 & 1 & 1 & & \\ \vdots & \vdots & \vdots & \ddots & \\ 1 & 1 & 1 & \ldots & 1 \end{bmatrix} \tag{5.5}$$

which contains a discrete, 0-degree (constant) spline in each row, with a knot placed at positions $1, 2, \ldots N$ respectively. This demonstrates that total variation denoising is also a *LASSO regression* problem using a set of constant splines as basis functions, and the aim is to produce a *sparse approximation* with as few non-zero knot coefficients as possible (Steidl et al., 2006; Kim et al., 2009).

The general LASSO regression problem has been studied extensively in the statistics and machine learning literature, and there are a large number of solvers that can be used to find the only minimum of the functional above. These include *subgradient* techniques such as *Gauss-Seidel* and *grafting* (Schmidt et al., 2007), but also methods that use a smoothed approximation to the 1-norm including *EpsL1*, *log-barrier*, *SmoothL1*, and *expectation-maximization* (Schmidt et al., 2007). Reformulation as a constrained least-squares problem leads to *interior-point*, *sequential quadratic programming* and variants (Schmidt et al., 2007). However, computational savings might be made by exploiting the special structure of this total variation regularization problem.

Minimizing the generalized functional with respect to variation in one of the variables $m_i$ alone (when the others are held fixed), can sometimes be conducted analytically, or is simple to compute approximately. This observation has lead to a number of very simple *coordinate descent* solvers for regularization problems (Friedman et al., 2007; Schmidt et al., 2007). It has been shown that such coordinate descent solvers are minimizers for functionals of the form:

$$H[m] = F[m, x] + \sum_{i=1}^{N} G_i(m_i) \tag{5.6}$$

where the likelihood functional term on the left is convex and differentiable, and the regularization functions $G_i$ are convex. The regularization term displayed here is *separable*: but the functionals in this paper do not have separable regularization terms. Special adaptations are therefore required in order to apply coordinate descent to the total variation regularization problem, for example see Friedman *et al.* (2007). This involves identifying the conditions under which groups of variables need to be merged and varied together. However, we make the observation here that the LASSO spline regression problem obtained from the total variation regularization method *is* separable, and that the spline regression matrix $S$ above has a particularly simple form. This allows us to develop a simple coordinate descent solver for total variation regularization that avoids the complexity of detecting and grouping variables altogether.

In particular, note that the original variables are obtained using $m = S\mu$ where $\mu$ are the spline coefficients, so that each element is just the cumulative sum of the spline coefficients:

$$m_i = \sum_{j=1}^{i} \mu_j \tag{5.7}$$

Similarly, going the other way, the spline coefficients can be obtained from the original variables using successive differences:

$$\mu_i = m_i - m_{i-1} \tag{5.8}$$

with $\mu_1 = m_1$. Also, note that at $\gamma = 0$, the original variables are equal to the input signal $x$, therefore the descent algorithm can be usefully initialised with the successive differences of the input signal. It is useful to understand this descent algorithm as a two-step process, (1) an *update step*:

$$w = \mu^k + S^{\star T}(x - S^\star \mu^k) \tag{5.9}$$

followed by (2) the *shrinkage step*:

$$\mu_i^{k+1} = \text{sgn}(w_i) \max(|w_i| - \gamma/\|S\|, 0) \tag{5.10}$$

with initial conditions $\mu_i^0 = (x_i - x_{i-1})\|S\|$, $\mu_1^0 = x_1\|S\|$, and $S^\star = S/\|S\|$, where $\|S\| = \sqrt{N(N+1)/2}$. Normalization of the spline matrix is required to prevent iterates from diverging. These steps (1), (2) are repeated until convergence. The original variables at convergence can be recovered using $m_i = 1/\|S\| \sum_{j=1}^{i} \mu_j$. We can understand (5.9) followed by (5.10) as the regression coefficient obtained by regressing the error $x - S^\star \mu^k$ in (5.9) onto the $i$-th variable $\mu_i^k$ (Friedman et al., 2007). The shrinkage term (5.10) is just the solution to the absolute penalized least-squares regression (5.4) if we fix all the variables except the variable $i$.

Using the observations about the matrix $S$ above, the update step can be simplified considerably:

$$w_i = \mu_i^k + \frac{1}{\|S\|} \sum_{j=i}^{N} \left( x_j - \frac{1}{\|S\|} \sum_{l=1}^{j} \mu_l^k \right) = \mu_i^k + \frac{1}{\|S\|} \sum_{j=i}^{N} x_j - \frac{1}{\|S\|^2} \sum_{j=i}^{N} \sum_{l=1}^{j} \mu_l^k \tag{5.11}$$

The expanded form of the expression on the right shows that the term in $x$ can be pre-computed, which can lead to further computational savings. Although simple, this coordinate descent algorithm requires a large number of iterations to reach convergence, particularly for small $\gamma$, because on each iteration, the variable $\mu_i$ does not change very much. Therefore, the speed of convergence is partly dependent upon the size of $\gamma$. Furthermore, the iterates before reaching convergence do not represent the solution at smaller values of $\gamma$, because $\gamma$ is fixed during the iteration. Thus, iteration of this algorithm does not obtain the *regularization path* automatically, as it does for the *piecewise linear regularization path follower* for the same problem (Hofling, 2009). For some applications, where we want the whole set of solutions when varying $\gamma \geq 0$, this could be a drawback. We note here that a related approach to PWC denoising was proposed independently in the geosciences literature (Mehta et al., 1990).

## 5.3 Robust total variation regularization and linear programming

Total variation regularization is a useful technique if the noise distribution is Gaussian. If there are outliers in the noise, then we can adapt the technique to increase its robustness by replacing the square likelihood loss with the absolute loss instead. The robust total variation functional becomes:

$$\Lambda = |x_i - m_j|I(i - j = 0) + \gamma|m_i - m_j|I(i - j = 1) \tag{5.12}$$

which can be cast as a *least absolute regression* problem:

$$m = \underset{m}{\operatorname{argmin}} H[m] = \underset{m}{\operatorname{argmin}} \left\| \begin{bmatrix} x \\ 0_N \end{bmatrix} - \begin{bmatrix} I_N \\ -\gamma \widetilde{D} \end{bmatrix} m \right\|_1 \tag{5.13}$$

where $I_N$ is the $N \times N$ identity matrix, $0_N$ is the $N \times 1$ zero matrix, $\|.\|_1$ is the vector 1-norm, and $\widetilde{D}$ is the $N \times N$ first difference matrix as in (4.26), but with the last row all zero. This is in the form of a *linear program* (a linear problem with linear inequality constraints), which is solvable using, for example, *simplex* or interior-point methods (Boyd and Vandenberghe, 2004; Koenker, 2005). To our knowledge though, specialized fast or regularization path-following methods for this robust total variation regularization problem do not exist, as they do for non-robust total variation regularization (but see Koenker et al. (1994) for related ideas, and Darbon and Sigelle (2006) for an approach in the case where the signals are integer rather than real, and also references therein).

## 5.4 Weighted convex clustering shrinkage

Convex clustering shrinkage has an advantage over mean shift and other clustering methods, that the functional is convex, so there exists a unique solution that minimizes the functional and it can be found by fast quadratic programming algorithms such as the interior point technique. However, the method can be highly sensitive to the choice of regularization parameter $\gamma$: there is typically only a small range over which the solution transitions from every sample belonging to its own cluster, to the emergence of a single cluster for all samples. To reduce this sensitivity and expand the useful range of the regularization parameter, a simple proposal is to focus the clustering only on those samples in $x$ that are initially close to each other. Samples that are far apart initially cannot therefore become clustered together. This leads to the following adaptation to the convex clustering shrinkage functional:

$$\Lambda = (1/2)|x_i - m_j|^2 I(i - j = 0) + \gamma|m_i - m_j|I(|x_i - x_j| \leq W) \tag{5.14}$$

This method retains the convexity properties of the original, because the weights are based on the input signal which is fixed. It is therefore amenable to quadratic programming. The parameter $W$ controls the extent of the value kernel, that is, how close the input samples need to be to be subject to sample distance reduction. As before, a small regularization parameter $\gamma$ constrains the solution to be similar to the input signal.

## 5.5 Convex mean shift clustering

The use of input-signal dependent weights for enhancing the usefulness of a PWC method presented above is a trick that can be applied more widely. For example, mean shift clustering is not convex, but it is possible to produce a simple adaptation that *is* convex:

$$\Lambda = |m_i - m_j|I(|x_i - x_j| \leq W) \tag{5.15}$$

for which the associated influence function is $I(|x_i - x_j| \leq W)\operatorname{sgn}(m_i - m_j)$. This should be contrasted with the influence function for mean shift clustering with absolute (rather than square) loss which is $I(|m_i - m_j| \leq W)\operatorname{sgn}(m_i - m_j)$. To see why this new method can be considered a convex version of mean shift clustering, consider that a solver for the descent ODEs for this method would be initialized with $m^0 = x$, such that, the influence function for the first iteration of this solver is $I(|m_i - m_j| \leq W)\operatorname{sgn}(m_i - m_j)$, and this coincides exactly with the influence function for (absolute) mean shift. The adaptive Euler solvers for the absolute mean shift and convex mean shift are, respectively:

$$m_i^{k+1} = m_i^k - \left( \sum_{j=1}^N I(|m_i^k - m_j^k| \leq W) \right)^{-1} \sum_{j=1}^N I(|m_i^k - m_j^k| \leq W)\operatorname{sgn}(m_i^k - m_j^k) \tag{5.16}$$

$$m_i^{k+1} = m_i^k - \left( \sum_{j=1}^N I(|x_i^k - x_j^k| \leq W) \right)^{-1} \sum_{j=1}^N I(|x_i^k - x_j^k| \leq W)\operatorname{sgn}(m_i^k - m_j^k) \tag{5.17}$$

(Note: with the square loss in classical mean shift in (4.16), the adaptive solver simplifies to the iterated mean, as shown earlier). One way of understanding the relationship to conventional mean shift is that the value kernel for convex mean shift does not change during iterations, whereas for mean shift the kernel weights are recomputed on each iteration.

### 5.6 Soft mean shift total variation diffusion and predictor-corrector integration

We have seen above (Section 4.6) that clustering methods have the PWC property in terms of level-sets, and total variation regularization in terms of splines. These different methods have certain disadvantages. The level-set representation is described in terms of levels, and this determines the locations of the jumps. A consequence of this is that rapid changes in the mean of the noise can cause rapid, spurious transitions between levels. On the other hand, the spline representation sets the location of the jumps, which in turn determines the constant levels. Therefore, the spline model is vulnerable to gradual, systematic changes in the level of constant regions due to changes in the mean of the noise, for example. Clustering methods such as mean shift provide constraints on the levels of constant regions and these could be used to alleviate the weaknesses of total variation algorithms, by contrast, the temporal constraints built into total variation algorithms could help prevent spurious transitions of clustering methods that are insensitive to temporal sequence.

Here we show that it is possible to synthesise the two representations using a novel PWC method that combines the global behaviour of mean shift clustering with the sequentially local behaviour of total variation regularization, using the following functional:

$$\Lambda = 1 - \exp\left(-\beta |x_i - m_j|^2/2\right)/\beta + \gamma |m_i - m_j| I(i-j=1) \quad (5.18)$$

Here, $\beta$ is a kernel parameter that determines the effective "precision" of the mean shift: if $\beta$ is large, then the solution can differentiate small peaks in the amplitude distribution, if small, then only large peaks are detected. Because of the form of (5.18), we call this method *soft mean shift total variation diffusion*. The regularization parameter $\gamma$ determines the relative influence of the total variation regularization term: if small, then locally sequential runs of close values have little influence over the solution; if large, then modes in the amplitude distribution can be broken up in order to find sequential constant runs instead.

Although not necessarily the best or most efficient solver, for the purposes of illustration we propose a two-step, *midpoint predictor-corrector* integrator for the resulting descent ODEs (Iserles, 2009):

$$m_i^\star = m_i^k - \frac{\Delta\eta}{2}\sum_{j=1}^{N} F'(m_i^k - x_j)\kappa_1(i-j) - \gamma\frac{\Delta\eta}{2}\sum_{j=1}^{K} G'(m_i^k - m_j^k)\kappa_2(i-j) \quad (5.19)$$

$$m_i^{k+1} = m_i^k - \Delta\eta \sum_{j=1}^{N} F'(m_i^\star - x_j)\kappa_1(i-j) - \gamma\Delta\eta \sum_{j=1}^{K} G'(m_i^\star - m_j^\star)\kappa_2(i-j) \quad (5.20)$$

with initial condition $m^0 = x$. Using this integrator, we obtain the following solver for this new PWC denoising algorithm:

$$m_i^\star = m_i^k - \frac{\Delta\eta}{2}\sum_{j=1}^{N} \exp\left(-\beta(m_i^k - x_j)^2/2\right)(m_i^k - x_j) - \gamma\frac{\Delta\eta}{2}\left[\text{sgn}(m_i^k - m_{i+1}^k) - \text{sgn}(m_i^k - m_{i-1}^k)\right] \quad (5.21)$$

$$\begin{aligned}m_i^{k+1} = m_i^k &- \Delta\eta \sum_{j=1}^{N} \exp\left(-\beta(m_i^\star - x_j)^2/2\right)(m_i^\star - x_j) \\ &- \gamma\Delta\eta[\text{sgn}(m_i^\star - m_{i+1}^\star) - \text{sgn}(m_i^\star - m_{i-1}^\star)]\end{aligned} \quad (5.22)$$

At the boundaries we have $m_i, m_i^\star \equiv 0$ for $i < 1$ and $i > N$ for the total variation part of the expression above. Although the regularization term is not differentiable everywhere, this finite difference solver is reasonably stable for small $\Delta\eta$, and experience shows that convergence to a useful, approximate solution is possible within a few hundred iterations.

## 6. Numerical results and discussion

In this section we discuss the results of applying the existing and new methods and solvers of this paper to typical PWC denoising problems. First, we focus on accurate recovery. We applied each method to a synthetic, unit step signal, corrupted by additive Gaussian noise. Method parameters were optimized by hand to achieve the output that is closest to the known step signal. We first tested the ability of the methods to recover the step whilst ignoring two isolated "outliers" that could be incorrectly identified as level transitions, the results are shown in Figure 3.

In the case of outliers, the new jump penalization and mean shift total variation diffusion methods (3k,l,j) appear to produce the most accurate results. Mean shift and bilateral filtering are able to recover the step (3d,h), but are unable to ignore the outliers. *K*-means can ignore the outliers (3f), but exhibits an incorrect transition near the leading step edge, because a sample near the edge is closer in value to the height of the step. Total variation regularization and the robust total variation regularization (3b,i) correctly ignore the outliers, but tend to identify many small, spurious edges; this is true also of iterated median filtering (3a). Although these spurious jumps in total variation methods can be removed by further increasing the regularization parameter, this will be at the expense of introducing very significant bias into the estimate of the level of the constant regions (essentially, this is a consequence of the piecewise linearity of the regularization path). There is, however, no corresponding parametric control over the iterated length 3 median filter, and clearly convergence on a root signal that has many spurious jumps. Soft mean shift, convex mean shift and weighted convex clustering shrinkage (3e,n,m) fail to ignore the outliers and also show some spurious transitions between levels. The objective step-fitting (see Table 2) method (3c) also places jumps at the outliers, and in other, spurious locations. Convex clustering shrinkage fails to identify the step at all and is also influenced by the outliers (3g).

Up until now, we have assumed that the noise is statistically independent, but in practice, it may have some kind of correlation. We therefore devised another challenging test: recover a unit step signal with linear drift in the mean of the noise as a confounding factor, see Figure 4. Now, we can see that mean shift, soft mean shift, *K*-means and mean shift total variation diffusion (4d,e,f,j) are able to recover the step and ignore the drift very effectively. These methods are successful in this case because they are largely insensitive to the sequential ordering of the input samples (with the exception of mean shift total variation diffusion); they are simply converging on peaks in the distribution of the input sample that turn out to be largely unaffected by the drift. Jump penalization, objective step-fitting, and bilateral methods (4k,l,c,h) are unable to ignore the drift, but produce the smoothest solutions. Weighted convex clustering shrinkage and convex mean shift (4m,n) are not confused by the drift, but have some spurious edges. Total variation regularization is also adversely affected by the drift and introduces a small, incorrect jump, but is appreciably better than robust total variation regularization (4b,i). Arguably the worst performing methods are iterated median filtering and convex clustering shrinkage (4a,g).

Next, in order to understand the efficiency of different methods and solvers, we apply a representative selection of iterative solvers to the basic task of noise removal from a short, unit synthetic step corrupted by Gaussian noise. Figure 5 shows the resulting output signal, and the *iteration path* of the solver: that is, the curves traced out by the samples in the solution as the iterations proceed. This is a plot of the iteration number on the horizontal axis, against the values of the samples $m_i$ on the vertical axis. The distance reduction principle is apparent in the output as the solver iterates towards convergence to a minimum of the associated functional. It is also possible to discriminate methods that use only value kernels such as mean shift and *K*-means, from methods that use local sequence kernels (for example, total variation regularization and bilateral filtering). The former can only merge together samples that are close in value, therefore, the iteration paths do not intersect. On the other hand, the latter can constrain those that are sequentially close to merge together, and the iteration paths can intersect.

In terms of the number of iterations, the forward stepwise jump placement algorithm for jump penalization methods are the most effective, converging on a solution in two steps (5f). Next, we find kernel adaptive step-size Euler integrators for mean shift, *K*-means and bilateral filtering taking at most five steps (5b,c,d). The forward linear regularization path following solver for total variation regularization is next, taking 10 steps to reach the unique optimum solution (5a). Weighted convex clustering shrinkage with non-adaptive step-size Euler integration takes some 300 steps to converge (5g). Lastly, the two-step mid-point predictor-corrector integrator for mean shift total variation regularization converges to a solution after about 500 iterations (5e).

Analytic minimizers for the generalized functional (3.1) are only available in the case of purely linear systems (simple quadratic loss functions). Therefore, numerical algorithms are required generally. The solvers described in this paper are not necessarily the most efficient that could be applied to each method. Comparing solvers is complicated and a fully rigorous approach is beyond the scope of this paper. However, there are some some general observations that can be made.

When the loss functions are convex and combined in convex combination this can be advantageous because then it is known that there is one unique minimizer for the functional, given fixed parameters. This avoids the uncertainty inherent to non-convex methods, where we do not know whether the solution obtained is the minimizer associated with the smallest possible value of the functional or not: there may be a better solution obtained by starting the solver from different initial conditions. This may require us to run the solver to convergence many times to gain confidence that the result is the best possible. Having said this, whether it matters that the solution is optimal depends on practical circumstances. For many PWC denoising methods the functionals are convex, and in terms of computational complexity, interior point algorithms are very efficient (Boyd and Vandenberghe, 2004).

If there are only a few jumps then forward stepwise jump placement, as described in Section 4.1, is very efficient. However, we cannot know whether a sequence of jumps placed by this forward-only algorithm is the best because the jump penalization functional is non-convex. Therefore, the same issues about uncertainty in the optimality of the results occur as with any non-convex functional. The scope for stepwise jump placement algorithms is quite narrow, because it requires an easily solvable likelihood function given the fixed spline knots.

Although having the widest scope of all, we have seen that finite difference methods for the descent ODEs can take hundreds of steps to converge, and are therefore relatively inefficient. However, the simple measure of adapting the step-size can cut the number of iterations required to reach convergence enormously, as we have seen for the mean shift and other clustering methods. Simple finite differences are only practical then if modified with adaptive step-sizes or some other approach to speeding up convergence.

The scope for (forward) piecewise linear regularization path followers for PWC denoising turns out to be reasonably wide (Rosset and Zhu, 2007), and if path linearity can be dropped, even wider (Rosset, 2004). Therefore, if the full regularization path of solutions is required, path following methods can be efficient, as we have seen for total variation regularization. To our knowledge, backwards path following has only been investigated for total variation regularization.

Coordinate descent is probably the least efficient in terms of number of iterations and requires separability of the regularization term, which does not apply in general to the PWC denoising functionals in this paper. However, the update on each iteration is very simple and this may yet turn out to be competitive with other solvers applied where separability can be shown to hold.

## 7. Summary, related and future ideas

In this paper, by presenting an extensively generalized mathematical framework for performing PWC noise removal, several new PWC denoising methods and associated solver algorithms are proposed that attempt to combine the advantages of existing methods in new and useful ways. Numerical tests on synthetic data compare the recovery accuracy and efficiency of these existing and novel methods head-to-head, finding that the new mean shift total variation denoising method is effective under challenging conditions where existing methods show significant deficiencies.

In order to devise these new PWC denoising methods, this study has presented a generalized approach to understanding and performing noise removal from piecewise constant signals. It is based on generalizing a substantial number of existing methods, found through a wide array of disciplines, under a generalized functional, where each method is associated with a special case of this functional. The generalized functional is constructed from all possible differences of samples in the input and output signals and their indices, over which simple and composite loss functions are placed. PWC outputs are obtained by seeking an output signal that minimizes the functional, which is a summation of these kernel loss functions. The task of PWC denoising is then formalized as the problem of recovering either a compact constant spline or level-set description of the PWC signal obscured by noise. Minimizing the functional is seen as constraining the difference between appropriate samples in the input signal. A range of solver algorithms for minimizing the functional are investigated, through which we were able to provide some novel observations on existing methods.

Whilst the structure of this paper has encouraged us to make as inclusive investigation as possible of PWC denoising methods, there are many other methods that cannot be associated with special cases of this generalized functional. Below, for completeness, we discuss the conceptual overlaps and relationships between some of these other methods that get significant use in practical PWC denoising applications.

### 7.1 Wavelets

Wavelet techniques are ubiquitous, generic methods for signal analysis, and their use in general noise removal has been comprehensively explored (Mallat and Hwang, 1992; Mallat and Zhong, 1992; Cattani, 2004; Mallat, 2009). Connections between wavelet techniques and some of the smoothing methods described in this paper, in particular total variation regularization (Steidl et al., 2004), have been established. Wavelet methods are powerful for many reasons, here we just mention a few of the basics: including (a) the existence of an algorithm with $O(N)$ computational complexity for the forward and reverse wavelet transforms in the discrete-time setting (Mallat, 2009); (b) the statistical theory of *wavelet shrinkage* that exploits orthonormality of the wavelet basis to perform noise removal using very simple, coefficient-by-coefficient (*separable*) nonlinear transformations of the wavelet coefficients (Candes, 2006); and (c) many signals, in the wavelet basis are *sparse*, that is, a large proportion of the coefficients are effectively zero making the wavelet representation very compact.

Wavelet methods require the choice of basis, and for PWC denoising, the *Haar basis*, itself composed of PWC functions, has been suggested many times in the wider literature (Cattani, 2004; Taylor et al., 2010), although it is not the only basis that has been proposed. Removing noise typically requires removal of the small-scale detail in the signal. The result of removing this detail is that the time-localisation of the remaining large scale PWC basis functions is poor, so that the jumps in the PWC signal cannot be accurately located and tend to become misaligned to the locations of the jumps in PWC bases instead. Furthermore, shrinkage causes "oscillations" near jumps that are not aligned with the jumps in the basis; oscillations that are similar in character to the Gibb's phenomena observed using linear low-pass filtering. These issues are an unavoidable consequence of the Heisenberg uncertainty inherent to time-frequency analysis (Mallat, 2009).

The PWC denoising methods described in this paper are not based on time-frequency analysis. Perhaps because of this, historically, wavelet-based approaches, and the kind of methods discussed in this paper, have developed quite separately (Candes and Guo, 2002). There are, however, some points of contact that have addressed how to prevent wavelet oscillations near jumps, yet retain some of the desirable conceptual and computational properties of wavelet methods. The literature on this topic is very extensive and we restrict ourselves to a few of the overlapping concepts that are of direct relevance to the PWC methods and solvers discussed in this paper.

If we are prepared to drop orthogonality, then we lose separability, but this does not mean that we lose the appealing concept of coefficient shrinkage: in fact, in the regression spline approach to total variation regularization discussed above, the use of the absolute function applied as a regularizer over the constant spline coefficients can be seen as *non-separable shrinkage* in the spline basis. The solver is more complex than separable shrinkage (we now have to solve a LASSO problem), but the jumps (spline knots) are no longer restricted by Heisenberg uncertainty and can be placed precisely at the jumps in the PWC signal (Mallat, 2009). Alternatively, Candes and Guo (2002) and others (Chan and Shen, 2005) discuss how the wavelet reconstruction with absolute loss on the wavelet coefficients can be augmented with the total variation of the wavelet reconstruction to attempt to minimize the oscillations near discontinuities. The solution can no longer be obtained using separable shrinkage, but the orthogonality and potential sparsity of the wavelet transform is retained. A final example is that of *iterated translation invariant wavelet shrinkage* (Steidl et al., 2004), which has been shown to have similar performance to total variation regularization, but the connection is somewhat less direct.

### 7.2 Hidden Markov models (HMMs)

*Hidden Markov models* (HMMs) play an important role in practical PWC denoising applications (Godfrey et al., 1980; Chung et al., 1990; Jong-Kae and Djuric, 1996; McKinney et al., 2006). It is important therefore to understand the relationships between the generalized methods proposed in this paper and HMMs. The literature on the very many variants of HMMs is extensive (Blimes, 2006), but we focus here on one of the most popular HMM variants that has seen repeated use in PWC denoising – the *discrete-state* HMM with *continuous, Gaussian emission probabilities*. This configuration has deep similarities to the (hard or soft) *K*-means clustering algorithms discussed in this paper. The similarity emerges from the relationship between *K*-means clustering and (Gaussian) *mixture density modelling*.

In this HMM variant, there are $K$ distinct *states* to the underlying *Markov chain*, each associated with a single Gaussian distribution, parameterised by $K$ means and variances. If the underlying chain is in state $s_i = k \in \{1,2 \dots K\}$ at index $i$, then the output sample from the noisy signal $x_i$ is drawn from a Gaussian with mean $\mu_k$ and variance $\sigma_k^2$. The Markov chain is parameterised by $K^2$ additional *transition density* and *initial probability* variables, the transition density determining the statistical dependence of $s_i$ upon $s_{i-1}$ and earlier states if necessary (Blimes, 2006).

The goal of fitting the HMM to the noisy signal is to find these transition and initial probabilities, and the parameters of the Gaussians associated with each state. If, however, $s_i$ is independent of $s_{i-1}$, then this HMM variant collapses to a Gaussian mixture density model (Roweis and Ghahramani, 1999), where the goal of fitting is to determine the parameters of the Gaussians alone. This is typically solved using *expectation-maximization* (EM) method (Hastie et al., 2001). There are two steps to this method, the *E-step*: in which the assignment of each index to each state is determined, and the *M-step* where the Gaussian parameters are re-estimated using the assignments. In this paper, the adaptive step-size Euler integrator applied to the *K*-means algorithms can be seen as a concatenation of these two steps, in the special case where the variances of the Gaussians are fixed. This arises because EM is equivalent to iterative, weighted mean and variance replacement, the weights determined by the state assignment. For soft *K*-means, the weights are the probabilities of assignment to each state given the means and variances from the previous iteration; for (hard) *K*-means, *most probable* assignments are used instead of probabilities, so the weights are either zero or one.

EM is has been adapted to the HMM case of mixture modelling, where $s_i$ depends on $s_{i-1}$. The E-step becomes more complex because calculating the state assignment probabilities requires "tracing" through all possible states up until index $i$. Fortunately, conditional independence of the Markov chain makes a considerable algebraic simplification of this assignment possible, in the probabilistic assignment case the resulting method is known as the *Baum-Welch* algorithm, the most probable variant of which is *Viterbi* or *sequential K-means training* (Blimes, 2006).

The means of the Gaussian associated with each state are analogous to the levels in the PWC level-set model, and this variant of HMM with continuous emission probabilities has the PWC property if the number of states $K \ll N$ because there will be many indices assigned sequentially to the same level. This explains why discrete-state HMMs with continuous emission probabilities are useful for general PWC denoising problems.

### 7.3 Piecewise constant (PWC) versus piecewise smooth (PWS)?

The fact that PWC signals are also *piecewise smooth* implies that methods for noise removal from piecewise smooth (PWS) signals can, in principle, be applied to the PWC denoising problem. Here, by PWS, we mean a signal that has a finite isolated set of discontinuities (jumps), and everywhere else the function has one or many continuous derivatives. The PWS noise removal problem has attracted considerable attention, in particular from those applying wavelet analysis in the signal and image processing communities (Chaisinthop and Dragotti, 2009; Mallat, 2009). For PWS signals, the level-set model is no longer parsimonious (but see the *stack* or *threshold decomposition* representation that is of central importance to morphological signal processing (Arce, 2005)). The extension of the 0-degree spline model to higher degrees requires piecewise (first, second etc.) differentiability, where the signal to be recovered is continuous everywhere, however, the PWC signals we refer to in this paper are discontinuous at the jump locations. Therefore, the higher degree spline model is not compact for PWS signals either. Here we discuss a small selection of PWS methods that are notable for their informative overlap with the algorithms in this paper.

Since noise removal from signals that are smooth everywhere is a problem for which the running mean filter is well suited, adapting the running linear filter to the existence of a few isolated jumps is a natural solution in many contexts. This requires some technique for (either implicitly or explicitly) detecting the existence of a jump. Many algorithms that provide jump capability to running filters (not just the running mean filter) exploit the concept of *data-adaptive weighting*, that is, some measure of the distance associated with samples inside (or outside) the local filtering window is used to provide a measure of whether a discontinuity exists within the window. This measure then changes the local weighting to mitigate the edge smoothing effect of filtering over the jump. In this paper, those techniques that place a kernel over the term $x_i - x_j$ are using such data-adaptive weighting.

In this context, it is informative to note that in the limit when $\beta \to 0$ in the bilateral filter formula, we obtain the iterated, running mean filter of width $W$, and with a soft (Gaussian) sequence kernel, we obtain the iterated running weighted mean filter. Therefore, one iteration of the bilateral filter can be viewed as a running weighted mean filter, where the weights are chosen to filter only those samples that are close in value (Elad, 2002). Similar ideas have been proposed independently in many different disciplines. Chung and Kennedy (1991) describe a weighted running mean filter with a weighting scheme that is constant but different in the left and right side of the window around each sample. The weights are inversely proportional to a positive power of the magnitude of the difference between the mean of the left or right sides of the window, and the sample in the middle of the window. The weights can be computed based on samples outside the filtering window, and the final output of the filter can be a summation over running means of differing lengths (Chung and Kennedy, 1991). Running filters based on a variety of linear combinations of rank ordered samples in the window, such as the *trimmed mean filter* or the *double window modified trimmed mean filter* are conceptually similar and very useful for PWS noise removal (Gather et al., 2006).

The PWC denoising algorithms in this paper are therefore closely related to some PWS algorithms, but the PWC denoising problem is distinct. In particular, we present evidence here that the PWC denoising problem is one for which information across the whole signal can be efficiently exploited by constructing a compact level-set representation, for example using the full pairwise differences in sample values in the mean shift or weighted convex clustering shrinkage algorithms. This approach would not be efficient for PWS signals, because in between the jumps, a PWS signal is not generally constant, and so does not necessarily have a compact level-set description.

### 7.4 Continuum approaches and nonlinear partial differential equations (PDEs)

The generalized functional (3.1) in this paper is based on a purely discrete-time setting. Most real signals are continuous in time, but despite continuous time being computationally inaccessible (it usually is), there are some mathematical advantages to going to a continuous time model of the signal, even if this has to be discretised later for computational reasons. The largest single class of continuous-time PWC denoising methods are those based on *nonlinear partial differential equations* (PDEs), and have nearly all been developed in the image processing literature (Chan and Shen, 2005). In the limit of infinitesimal time increments, the discrete-time, generalized functional becomes a double integral functional instead. Then, the *variational derivative* of the functional with respect to the continuous-time output signal is an *Euler-Lagrange* PDE, and it will be nonlinear if it is useful for PWC denoising. So, it is fairly easy to show that many, if not most, of the methods in this paper have an equivalent PDE form. Numerical solvers for this PDE would be very similar to numerical solvers for the descent ODEs derived earlier. Passing to the continuum also invites application of *Sethian's computational level-set algorithms* that, in the 1D signal case, would correspond to techniques for evolving the jump locations between the distinct level-sets that comprise the PWC solution, as opposed to the levels (Chan and Shen, 2005).

### 7.5 Future directions

The new methods and solvers presented in this paper represent just a handful of directions that the generalized functional and solver description suggests. Clearly, there are a very large number of other possible methods that can be constructed from the functional components we describe in this paper, that are as yet unexplored, that might be of value in PWC denoising. However, determining which of these methods would have minimizer(s) with the PWC property, and in addition, admit efficient and reliable solvers, will require additional work. We imagine one approach: a formal axiomatic system leading to the *scale-space equation* has been developed to the design of nonlinear PDEs for image analysis, that constrains their form to have universally useful properties (Chan and Shen, 2005). It is quite possible that such axioms might be modified for PWC denoising purposes. The consequences of such axioms could be explored with respect to the functional components and their interactions with the solvers presented in this paper, with a view to asking what combinations lead to solutions with the PWC property.

## Appendix

To prove that the 3-point iterated median filter cannot raise the total variation of the signal, we examine two adjacent windows and apply a simple combinatorial argument over the input signal $x_1, x_2, x_3, x_4$, so that the two input windows have the values $x_2, x_3$, and the two output windows have the values $y_2 = \text{median}(x_1, x_2, x_3)$ and $y_3 = \text{median}(x_2, x_3, x_4)$. Now, label $x_2, x_3$ as "inner" values, and the other two as "outer" values. The non-increasing total variation condition is that $|y_2 - y_3| \leq |x_2 - x_3|$. Since the median operation selects one of the values in the input set, there are four different cases to consider. First, consider when both windows select the same input, i.e. $y_2 = y_3$, their difference is zero and the condition is satisfied trivially. Similarly trivial is the case when the two inner values are swapped, i.e. $y_2 = x_3$ and $y_3 = x_2$, the condition is satisfied at equality. Thirdly, if one of the windows selects one of the inner values, and the other one of the outer values, then it must be that the selected outer value lies in between the two inner values, and so is closer to either of the inner values than the inner values are to themselves, satisfying the condition. The final case is when both outer values $x_1, x_4$ are selected, but in that case they both lie in between the inner values and so the condition is again satisfied. This proves that $|y_2 - y_3| \leq |x_2 - x_3|$ implying that the median operation applied to these two windows cannot increase the total variation. The final step in the proof is to extend this to the entire signal: the total variation over every pair of adjacent values cannot increase, so the total variation over the entire signal cannot increase either. Thus, 3-point median filtering can only either leave the total variation of a signal unchanged or reduce it after each iteration.

### References

Arce GR (2005) Nonlinear signal processing: a statistical approach. Hoboken, N.J.: Wiley-Interscience.
Arias-Castro E, Donoho DL (2009) Does Median Filtering Truly Preserve Edges Better Than Linear Filtering? Annals of Statistics 37:1172-1206.
Blimes JA (2006) What HMMs can do. IEICE Transactions on Information and Systems E89-D:869-891.
Bloom HJ (2009) Next generation Geostationary Operational Environmental Satellite: GOES-R, the United States' advanced weather sentinel. In: Remote Sensing System Engineering II, 1 Edition, pp 745802-745809. San Diego, CA, USA: SPIE.
Boyd SP, Vandenberghe L (2004) Convex optimization. Cambridge, UK ; New York: Cambridge University Press.
Candes EJ (2006) Modern statistical estimation via oracle inequalities. Acta Numerica 15:257-326.
Candes EJ, Guo F (2002) New multiscale transforms, minimum total variation synthesis: applications to edge-preserving image reconstruction. Signal Processing 82:1519-1543.
Cattani C (2004) Haar wavelet-based technique for sharp jumps classification. Mathematical and Computer Modelling 39:255-278.

Chaisinthop V, Dragotti PL (2009) Semi-parametric compression of piecewise-smooth functions. In: Proceedings of the European Conference on Signal Processing (EUSIPCO). Glasgow, UK.

Chan TF, Shen J (2005) Image processing and analysis : variational, PDE, wavelet, and stochastic methods. Philadelphia: Society for Industrial and Applied Mathematics.

Cheng YZ (1995) Mean Shift, Mode Seeking, and Clustering. IEEE Transactions on Pattern Analysis and Machine Intelligence 17:790-799.

Chung SH, Kennedy RA (1991) Forward-Backward Nonlinear Filtering Technique for Extracting Small Biological Signals from Noise. Journal of Neuroscience Methods 40:71-86.

Chung SH, Moore JB, Xia L, Premkumar LS, Gage PW (1990) Characterization of Single Channel Currents Using Digital Signal-Processing Techniques Based on Hidden Markov-Models. Philosophical Transactions of the Royal Society of London Series B-Biological Sciences 329:265-285.

Darbon J, Sigelle M (2006) Image restoration with discrete constrained total variation - Part I: Fast and exact optimization. Journal of Mathematical Imaging and Vision 26:261-276.

Dong YQ, Chan RH, Xu SF (2007) A detection statistic for random-valued impulse noise. IEEE Transactions on Image Processing 16:1112-1120.

Elad M (2002) On the origin of the bilateral filter and ways to improve it. IEEE Transactions on Image Processing 11:1141-1151.

Fried R (2007) On the robust detection of edges in time series filtering. Computational Statistics & Data Analysis 52:1063-1074.

Friedman J, Hastie T, Hofling H, Tibshirani R (2007) Pathwise Coordinate Optimization. Annals of Applied Statistics 1:302-332.

Fukunaga K, Hostetler L (1975) The estimation of the gradient of a density function, with applications in pattern recognition. IEEE Transactions on Information Theory 21:32-40.

Gather U, Fried R, Lanius V (2006) Robust detail-preserving signal extraction. In: Handbook of time series analysis (Schelter B, Winterhalder M, Timmer J, eds), pp 131-153: Wiley-VCH.

Gill D (1970) Application of a Statistical Zonation Method to Reservoir Evaluation and Digitized-Log Analysis. American Association of Petroleum Geologists Bulletin 54:719-&.

Godfrey R, Muir F, Rocca F (1980) Modeling Seismic Impedance with Markov-Chains. Geophysics 45:1351-1372.

Hastie T, Tibshirani R, Friedman JH (2001) The elements of statistical learning : data mining, inference, and prediction : with 200 full-color illustrations. New York: Springer.

Hofling H (2009) A path algorithm for the fused lasso signal approximator. In: arXiV.

Iserles A (2009) A first course in the numerical analysis of differential equations, 2nd Edition. Cambridge ; New York: Cambridge University Press.

Jong-Kae F, Djuric PM (1996) Automatic segmentation of piecewise constant signal by hidden Markov models. In: Statistical Signal and Array Processing, 1996. Proceedings., 8th IEEE Signal Processing Workshop on (Cat. No.96TB10004, pp 283-286.

Justusson BI (1981) Median filtering: statistical properties. In: Two-Dimensional Digital Signal Prcessing II: Transforms and Median Filters (Huang TS, ed), pp 161-196: Springer.

Kalafut B, Visscher K (2008) An objective, model-independent method for detection of non-uniform steps in noisy signals. Computer Physics Communications 179:716-723.

Kerssemakers JWJ, Munteanu EL, Laan L, Noetzel TL, Janson ME, Dogterom M (2006) Assembly dynamics of microtubules at molecular resolution. Nature 442:709-712.

Kim SJ, Koh K, Boyd S, Gorinevsky D (2009) L1 Trend Filtering. SIAM Review 51:339-360.

Koenker R (2005) Quantile regression. Cambridge ; New York: Cambridge University Press.

Koenker R, Ng P, Portnoy S (1994) Quantile Smoothing Splines. Biometrika 81:673-680.

Mallat S, Hwang WL (1992) Singularity Detection and Processing with Wavelets. IEEE Transactions on Information Theory 38:617-643.

Mallat S, Zhong S (1992) Characterization of Signals from Multiscale Edges. IEEE Transactions on Pattern Analysis and Machine Intelligence 14:710-732.

Mallat SG (2009) A wavelet tour of signal processing : the sparse way, 3rd Edition. Amsterdam ; Boston: Elsevier/Academic Press.

Marr D, Hildreth E (1980) Theory of Edge-Detection. Proceedings of the Royal Society of London Series B-Biological Sciences 207:187-217.

McKinney SA, Joo C, Ha T (2006) Analysis of single-molecule FRET trajectories using hidden Markov modeling. Biophysical Journal 91:1941-1951.

Mehta CH, Radhakrishnan S, Srikanth G (1990) Segmentation of Well Logs by Maximum-Likelihood-Estimation. Mathematical Geology 22:853-869.

Mrazek P, Weickert J, Bruhn A (2006) On Robust Estimation and Smoothing with Spatial and Tonal Kernels. In: Geometric Properties for Incomplete Data (Klette R, Kozera R, Noakes L, Weickert J, eds): Springer.

OLoughlin KF (1997) SPIDR on the Web: Space physics interactive data resource on-line analysis tool. Radio Science 32:2021-2026.

Page ES (1955) A Test for a Change in a Parameter Occurring at an Unknown Point. Biometrika 42:523-527.

Papadimitriou CH, Steiglitz K (1998) Combinatorial optimization : algorithms and complexity. Mineola, N.Y.: Dover Publications.

Pawlak M, Rafajlowicz E, Steland A (2004) On detecting jumps in time series: Nonparametric setting. Journal of Nonparametric Statistics 16:329-347.

Pelckmans K, de Brabanter J, Suykens JAK, de Moor B (2005) Convex Clustering Shrinkage. In: PASCAL Workshop on Statistics and Optimization of Clustering.

Perona P, Malik J (1990) Scale-Space and Edge-Detection Using Anisotropic Diffusion. IEEE Transactions on Pattern Analysis and Machine Intelligence 12:629-639.

Pilizota T, Brown MT, Leake MC, Branch RW, Berry RM, Armitage JP (2009) A molecular brake, not a clutch, stops the Rhodobacter sphaeroides flagellar motor. Proceedings of the National Academy of Sciences of the United States of America 106:11582-11587.

Rosset S (2004) Tracking curved regularization optimization solution paths. In: Advances in Neural Information Processing: MIT Press.

Rosset S, Zhu J (2007) Piecewise linear regularized solution paths. Annals of Statistics 35:1012-1030.

Roweis S, Ghahramani Z (1999) A unifying review of linear gaussian models. Neural Computation 11:305-345.

Rudin LI, Osher S, Fatemi E (1992) Nonlinear Total Variation Based Noise Removal Algorithms. Physica D 60:259-268.

Ryder RT, Crangle RD, Trippi MH, Swezey CS, Lentz EE, Rowan LR, Hope RS (2009) Geologic Cross Section D–D' Through the Appalachian Basin from the Findlay Arch, Sandusky County, Ohio, to the Valley and Ridge Province, Hardy County, West Virginia. In: U.S. Geological Survey Scientific Investigations, p 52: U.S. Geological Survey.

Schmidt M, Fung G, Rosales R (2007) Fast optimization methods for L1 regularization: A comparative study and two new approaches. Machine Learning: ECML 2007, Proceedings 4701:286-297
809.

Serra JP (1982) Image analysis and mathematical morphology. London ; New York: Academic Press.

Snijders AM, Nowak N, Segraves R, Blackwood S, Brown N, Conroy J, Hamilton G, Hindle AK, Huey B, Kimura K, Law S, Myambo K, Palmer J, Ylstra B, Yue JP, Gray JW, Jain AN, Pinkel D, Albertson DG (2001) Assembly of microarrays for genome-wide measurement of DNA copy number. Nature Genetics 29:263-264.

Sowa Y, Rowe AD, Leake MC, Yakushi T, Homma M, Ishijima A, Berry RM (2005) Direct observation of steps in rotation of the bacterial flagellar motor. Nature 437:916-919.

Steidl G, Didas S, Neumann J (2006) Splines in higher order TV regularization. International Journal of Computer Vision 70:241-255.

Steidl G, Weickert J, Brox T, Mrazek P, Welk M (2004) On the equivalence of soft wavelet shrinkage, total variation diffusion, total variation regularization, and SIDEs. Siam Journal on Numerical Analysis 42:686-713.

Strong D, Chan T (2003) Edge-preserving and scale-dependent properties of total variation regularization. Inverse Problems 19:S165-S187.

Taylor JN, Makarov DE, Landes CF (2010) Denoising Single-Molecule FRET Trajectories with Wavelets and Bayesian Inference. Biophysical journal 98:164-173.

Tibshirani RJ, Taylor J (2010) Regularization paths for least squares problems with generalized L1 penalties. In: arXiv.

Tukey JW (1977) Exploratory data analysis. Reading, Mass.: Addison-Wesley Pub. Co.

Table 1: "Components" for PWC denoising methods. All the methods in this paper can be constructed using all pairwise differences between input samples, output samples, and sequence indices. These differences are then used to define kernel and loss functions. Loss functions and kernels are combined to make the generalized functional to be minimized with respect to the output signal $m$. Function $I(S)$ is the indicator function such that $I(S) = 1$ if the condition $S$ is true, and $I(S) = 0$ otherwise.

| (a) Difference $d$ | Description |
|---|---|
| $x_i - m_j$ | Input-output value difference; used in likelihood terms |
| $m_i - m_j$ | Output-output value difference; used in regularization terms |
| $x_i - x_j$ | Input-input value difference; used in both likelihood and regularization terms |
| $i - j$ | Sequence difference; used in both likelihood and regularization terms |

| (b) Kernel function | Description |
|---|---|
| 1 | Global |
| $I(|d| \leq W)$ $I(|d|^2/2 \leq W)$ | Hard (local in either value or sequence) |
| $\exp(-\beta|d|)$ $\exp(-\beta|d|^2/2)$ | Soft (semi-local in either value or sequence) |
| $I(d = 1)$ | Isolates only sequentially adjacent terms when used as sequence kernel |
| $I(d = 0)$ | Isolates only terms that have the same index when used as sequence kernel |

| (c) Loss function | Influence function (derivative of loss function) $kernel \times direction$ | Composition |
|---|---|---|
| $L_0(d) = |d|^0$ $L_1(d) = |d|^1$ $L_2(d) = |d|^2/2$ | $L'_1(d) = 1 \times \text{sgn}(d)$ $L'_2(d) = 1 \times d$ | Simple |
| $L_{W,1}(d) = \min(|d|, W)$ $L_{W,2}(d) = \min(|d|^2/2, W)$ | $L'_{W,1}(d) = I(|d| \leq W) \times \text{sgn}(d)$ $L'_{W,2}(d) = I(|d|^2/2 \leq W) \times d$ | Composite |
| $L_{\beta,1}(d) = 1 - \exp(-\beta|d|)/\beta$ $L_{\beta,2}(d) = 1 - \exp(-\beta|d|^2/2)/\beta$ | $L'_{\beta,1}(d) = \exp(-\beta|d|) \times \text{sgn}(d)$ $L'_{\beta,2}(d) = \exp(-\beta|d|^2/2) \times d$ | Composite |

Table 2: A generalized functional for noise removal from piecewise constant (PWC) signals. The functional combines differences, losses and kernel functions described in Table 1 into a function to be minimized over all samples, pairwise. Various solver algorithms are used to minimize this functional with respect to the solution $m$, these are described in Table 3.

| **Generalized functional for piecewise constant noise removal** |||
|---|---|---|
| $$H[m] = \sum_{i=1}^{N}\sum_{j=1}^{N} \Lambda(x_i - m_j, m_i - m_j, x_i - x_j, i - j)$$ |||
| **Existing methods** | **Function $\Lambda$** | **Notes** |
| Linear diffusion | $(1/2)\|m_i - m_j\|^2 I(i - j = 1)$ | *Solved by weighted mean filtering; cannot produce PWC solutions; not PWC* |
| Step-fitting (Gill, 1970; Kerssemakers et al., 2006) | $(1/2)\|x_i - m_j\|^2 I(i - j = 0)$ | *Termination criteria based on number of jumps; PWC* |
| Objective step-fitting (Kalafut and Visscher, 2008) | $(1/2)\|x_i - m_j\|^2 I(i - j = 0) + \lambda\|m_i - m_j\|^0 I(i - j = 1)$ | *Likelihood term the same upto log transformation; regularization parameter $\lambda$ fixed by data; PWC* |
| Total variation regularization (Rudin et al., 1992) | $(1/2)\|x_i - m_j\|^2 I(i - j = 0) + \gamma\|m_i - m_j\| I(i - j = 1)$ | *Convex; fused LASSO signal approximator is the same; PWC* |
| Total variation diffusion | $\|m_i - m_j\| I(i - j = 1)$ | *Convex; partially minimized by iterated 3-point median filter; PWC* |
| Mean shift clustering | $\min\left((1/2)\|m_i - m_j\|^2, W\right)$ | *Non-convex; PWC* |
| Likelihood mean shift clustering | $\min\left((1/2)\|x_i - m_j\|^2, W\right)$ | *Non-convex; K-means is similar but not a direct special case (see text); PWC* |
| Soft mean shift clustering | $1 - \exp\left(-\beta\|m_i - m_j\|^2/2\right)/\beta$ | *Non-convex; PWC* |
| Soft likelihood mean shift clustering | $1 - \exp\left(-\beta\|x_i - m_j\|^2/2\right)/\beta$ | *Non-convex; soft-K-means is similar but not a direct special case (see text); PWC* |
| Convex clustering shrinkage (Pelckmans et al., 2005) | $(1/2)\|x_i - m_j\|^2 I(i - j = 0) + \gamma\|m_i - m_j\|$ | *Convex; PWC* |
| Bilateral filter (Mrazek et al., 2006) | $\left[1 - \exp\left(-\beta\|m_i - m_j\|^2/2\right)/\beta\right] I(\|i - j\| \leq W)$ | *Non-convex* |
| **New methods proposed in this paper** |||
| Jump penalization | $(1/2)\|x_i - m_j\|^2 I(i - j = 0) + \gamma\|m_i - m_j\|^0 I(i - j = 1)$ | *Non-convex; PWC* |
| Robust jump penalization | $\|x_i - m_j\| I(i - j = 0) + \gamma\|m_i - m_j\|^0 I(i - j = 1)$ | *Non-convex; PWC* |
| Robust total variation regularization | $\|x_i - m_j\| I(i - j = 0) + \gamma\|m_i - m_j\| I(i - j = 1)$ | *Convex; PWC* |
| Soft mean shift total variation diffusion | $1 - \exp\left(-\beta\|x_i - m_j\|^2/2\right)/\beta + \gamma\|m_i - m_j\| I(i - j = 1)$ | *Non-convex; PWC* |
| Weighted convex clustering shrinkage | $(1/2)\|x_i - m_j\|^2 I(i - j = 0) + \gamma\|m_i - m_j\| \exp(-\beta\|x_i - x_j\|)$ | *Convex; PWC* |
| Convex mean shift clustering | $\|m_i - m_j\| \exp(-\beta\|x_i - x_j\|)$ | *Convex; PWC* |

Table 3: Solvers for finding a minimizer of the generalized piecewise constant (PWC) noise removal functional in Table 2. The first column is the solver algorithm, the second is the different PWC methods to which the technique can be applied in theory.

| Solver | Can apply to | Notes |
|---|---|---|
| Analytic convolution | Linear diffusion | *Problems with only square loss functions are analytical in a similar way* |
| Linear programming (Boyd and Vandenberghe, 2004) | Robust total variation regularization | *Direct minimizer of functional; also all piecewise linear convex problems* |
| Quadratic programming (Boyd and Vandenberghe, 2004) | Total variation regularization  Convex clustering shrinkage | *Direct minimizer of functional; also all problems that combine square likelihood with absolute regularization loss* |
| Stepwise jump placement (Gill, 1970; Kerssemakers et al., 2006; Kalafut and Visscher, 2008) | Step-fitting  Objective step-fitting  Jump penalization  Robust jump penalization | *Greedy spline fit minimizer of functional* |
| Finite differencing (Mrazek et al., 2006) | Total variation regularization  Total variation diffusion  Convex clustering shrinkage  Mean shift clustering  Likelihood mean shift clustering  Soft mean shift clustering  Soft *K*-means clustering  Robust total variation regularization  Soft mean shift total variation diffusion | *Finite differences are not guaranteed to converge for non-differentiable loss functions* |
| Coordinate descent (Friedman et al., 2007) | Total variation regularization  Robust total variation regularization | |
| Iterated mean replacement (Cheng, 1995) | Mean shift clustering  Likelihood mean shift clustering | *Obtainable as adaptive step-size forward Euler differencing* |
| Weighted iterated mean replacement (Cheng, 1995) | Soft mean shift clustering  Soft likelihood mean shift clustering | *Obtainable as adaptive step-size forward Euler differencing* |
| Piecewise linear regularization path follower (Rosset and Zhu, 2007; Hofling, 2009) | Total variation regularization  Convex clustering shrinkage | |
| Least-angle regression path follower (Tibshirani and Taylor, 2010) | Total variation regularization | *Reverse of piecewise linear regularization path follower* |

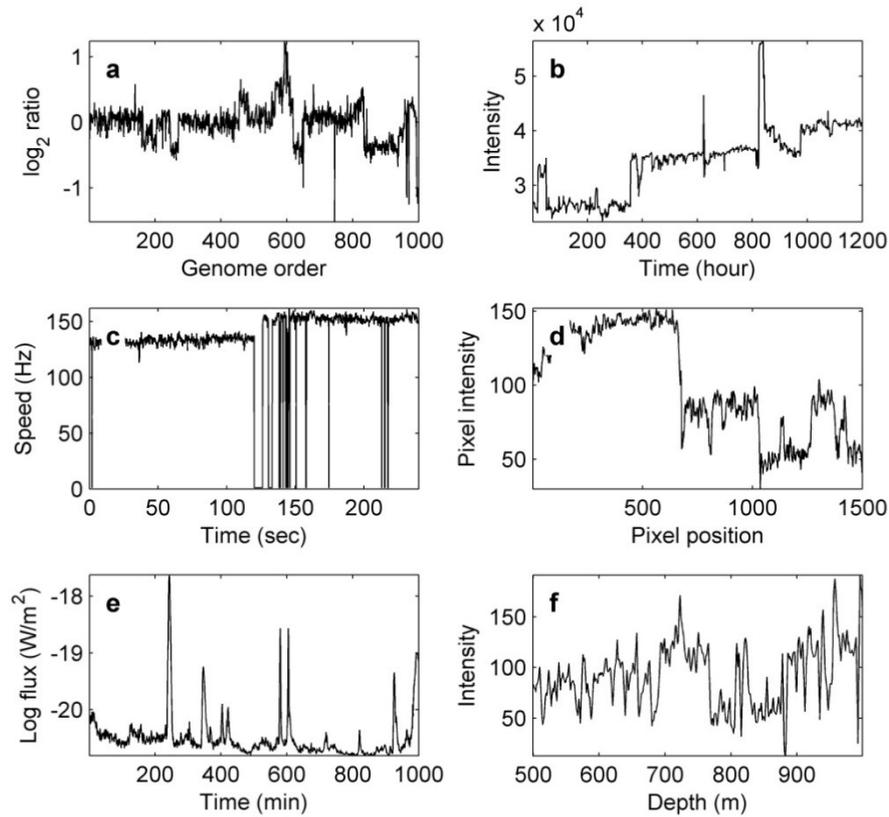

Figure 1: Examples of signals that could be modeled as piecewise constant (PWC) signals obscured by noise. (a) Log normalized DNA copy-number ratios against genome order from a microarray-based comparative genomic hybridization study (Snijders et al., 2001); (b) Cosmic ray intensity against time recorded by neutron monitor (OLoughlin, 1997); (c) rotation speed against time of R. Sphaeroides flagellum (Pilizota et al., 2009), (d) pixel red intensity value against horizontal pixel position for a single scan line from a digital image, (e) short-wavelength solar X-ray flux against time recorded by GOES-15 space weather satellite (Bloom, 2009), and (f) gamma ray intensity against depth from USGS wireline geological survey well log (Ryder et al., 2009).

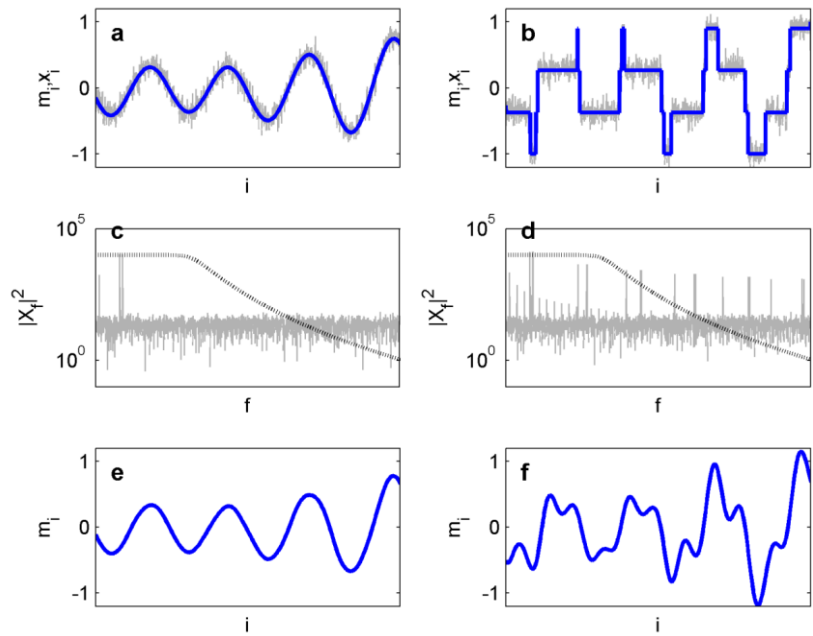

Figure 2: Noise removal from PWC signals is a task for which no linear filter is efficient, because, for independent noise, the noise and the PWC signal both have *infinite bandwidth*, e.g. there is no maximum frequency above which the Fourier components of either have zero magnitude. (a) A smooth signal (blue) with added noise (grey), constructed from a few sinusoidal components of random frequency and amplitude; (b) a PWC signal (blue) with added noise (grey), constructed from "square-wave" components of the same frequency and amplitude as the smooth signal. (c) (Discrete) Fourier analysis of the noisy smooth signal shows a few large magnitude, low-frequency components, and the background noise level that occupies the whole frequency range. (d) Fourier analysis of the noisy PWC signal in (b), showing the same low-frequency, large magnitude components, but also many other large magnitude components across the entire frequency range (which are harmonics of the square wave components). The black, dotted line in (c) and (d) shows the frequency response (magnitude not to scale) of a low-pass filter used to perform noise removal; this is applied to the noisy, smooth signal in (e) and the noisy PWC signal in (f). It can be seen that whilst the smooth signal is recovered effectively and there is little noticeable distortion, although noise is removed from the PWC signal, the jumps are also smoothed away or cause spurious oscillations (Gibb's phenomena).

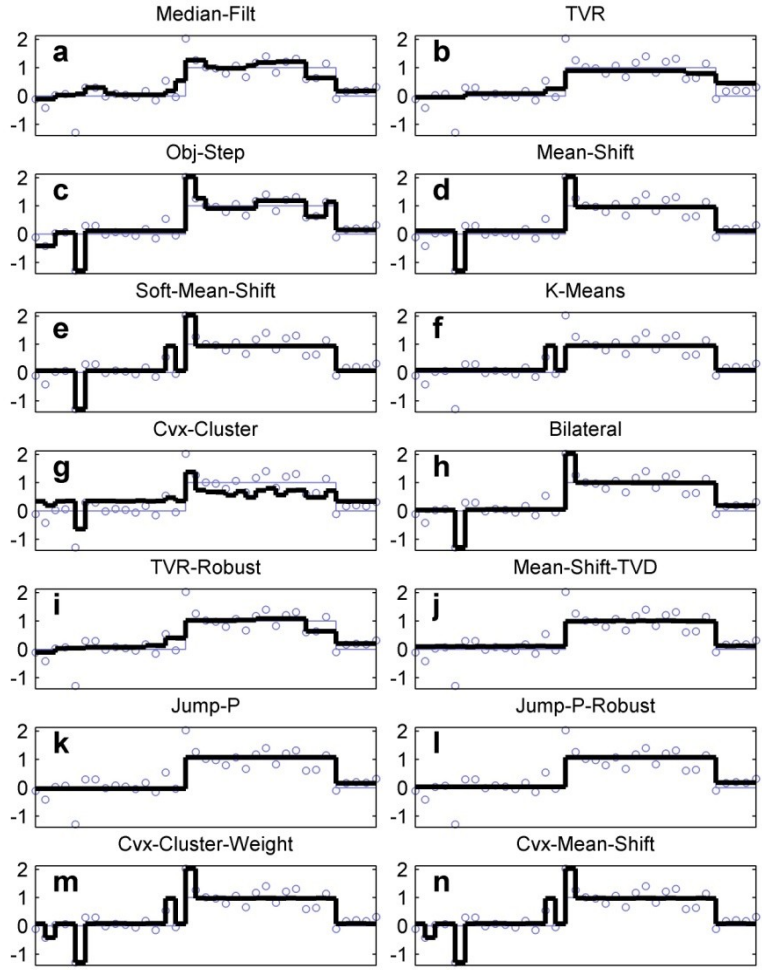

Figure 3: Response of PWC denoising methods to a step of unit height with additive Gaussian noise ($\sigma = 0.25$) and two extreme outliers. The methods are (a) iterated median filter for total variation diffusion, (b) total variation regularization ($\gamma = 1.5$), (c) objective step-fitting, (d) mean shift ($W = 0.42$), (e) soft mean shift ($\beta = 15$), (f) $K$-means ($K = 2$), (g) convex clustering shrinkage ($\gamma = 0.02$), (h) bilateral filter ($W = 2, \beta = 10$), (i) robust total variation regularization ($\gamma = 1.5$), (j) soft mean shift total variation diffusion ($\beta = 10, \gamma = 2.0$), (k) jump penalization ($\gamma = 1.0$), (l) robust jump penalization ($\gamma = 3.0$), (m) weighted convex clustering shrinkage ($\gamma = 1.0, W = 0.22$), and (n) convex mean shift ($\gamma = 1.0, W = 0.22$).

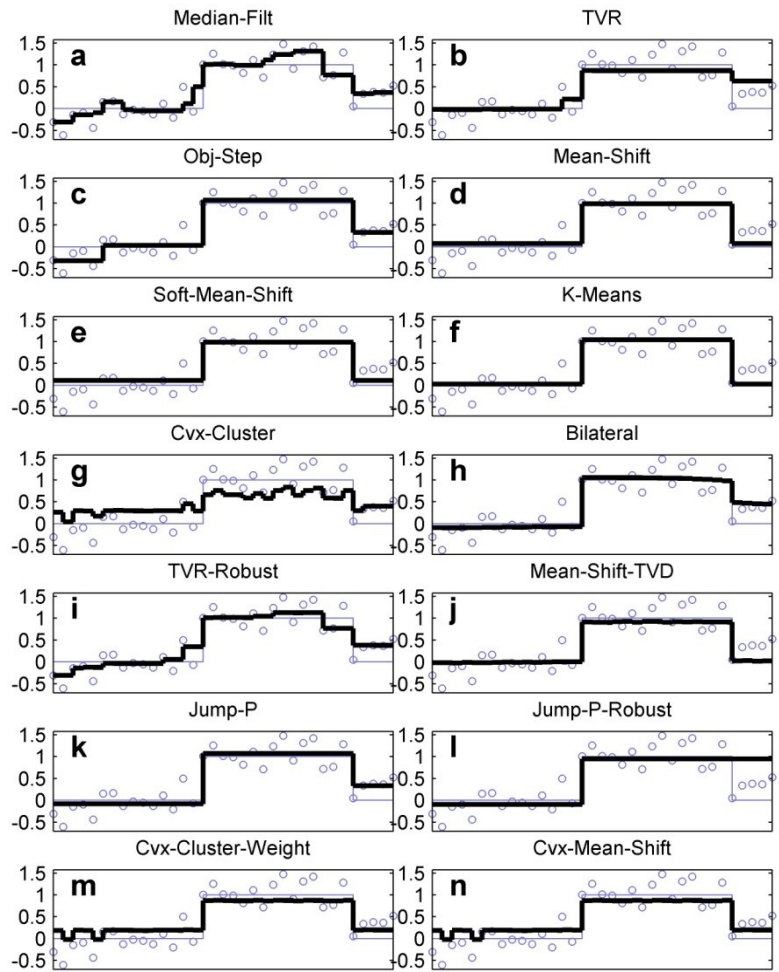

Figure 4: Response of PWC denoising methods to a step of unit height with additive Gaussian noise ($\sigma = 0.25$) and linear mean drift. The methods and parameters are as described in Figure 3.

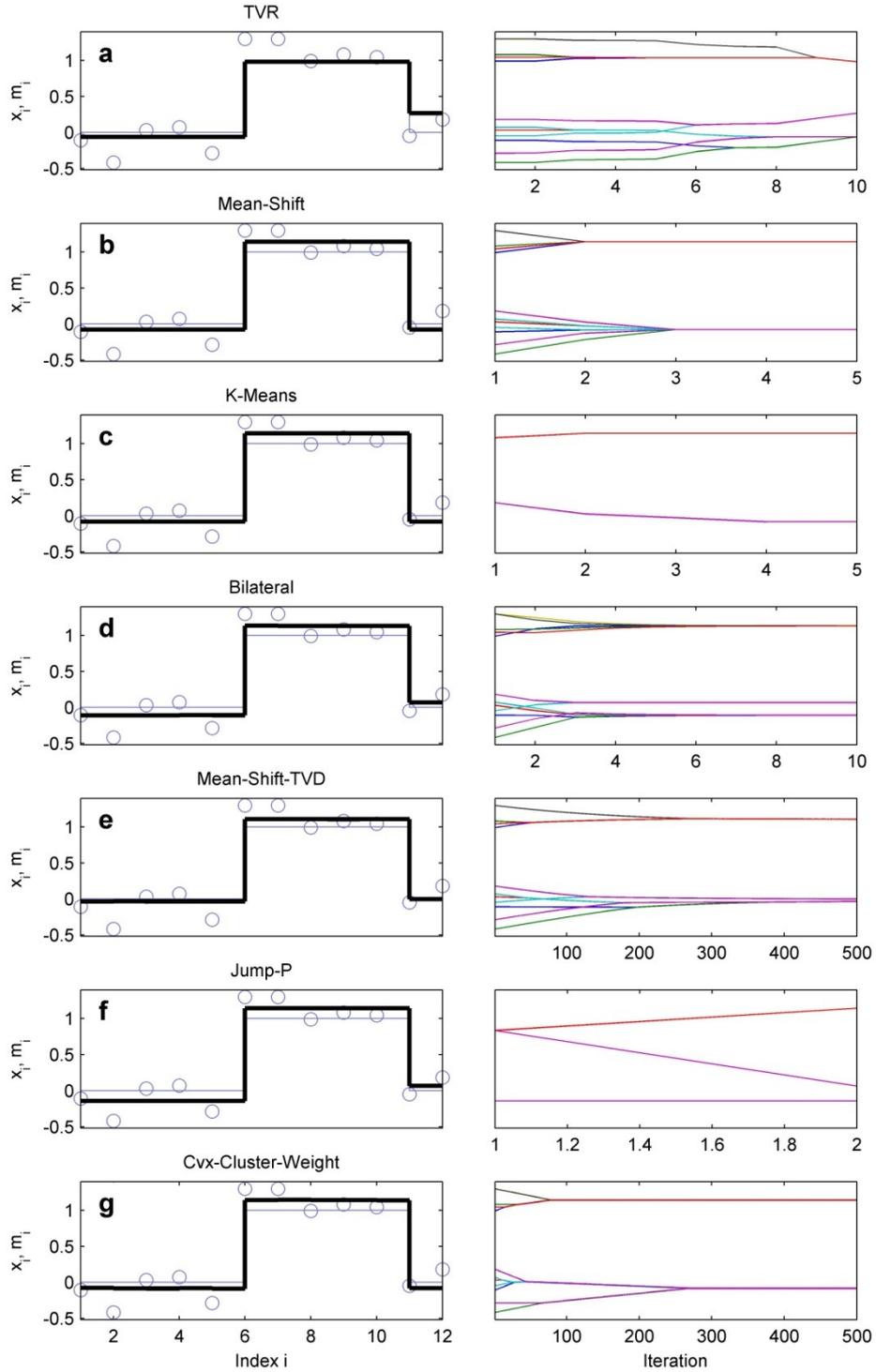

Figure 5: Iteration paths for solvers applied to a representative sample of PWC denoising methods. The noise Is Gaussian ($\sigma = 0.25$). The left column shows the final, converged outputs of each method, the right column the associated iteration path taken towards convergence. The vertical axes are the values of the input (blue circles) and output (black line) samples, and the known PWC signal (thin blue line). The methods and solver algorithms are (a) total variation regularization by piecewise linear forward regularization path follower, (b) mean shift with adaptive step-size Euler integration, (c) *K*-means with adaptive step-size Euler integration, (d) bilateral filtering with adaptive step-size Euler integration (e) mean shift total variation diffusion with predictor-corrector two-step integration, (f) jump penalization with forward stepwise jump placement, (g) weighted convex clustering shrinkage with Euler integration. Method parameters are chosen to give good PWC recovery results.